\documentclass[%
nofootinbib,
 amsmath,amssymb,
 aps,
 prd,
]{revtex4-1}

\usepackage{slashed}
\usepackage{graphicx}
\usepackage{dcolumn}
\usepackage{bm}
\usepackage{hyperref}

\usepackage{float}
\usepackage{longtable}

\newcommand{\forests}{\mathfrak{F}}
\newcommand{\forestsmax}{\mathfrak{F}_{\mathrm{max}}}

\newcommand{\mcmain}{\text{main}}
\newcommand{\mcmin}{\text{min}}
\newcommand{\mcuniform}{\text{uniform}}
\newcommand{\mcmodify}{\text{modify}}
\newcommand{\ffdeg}{\mathrm{Deg}}
\newcommand{\mcdegsub}{S^{\text{sub}}}
\newcommand{\mcdegsat}{S^{\text{sat}}}
\newcommand{\mcdegmul}{S^{\text{mul}}}

\newcommand{\nloops}{\mathrm{Loop}}
\newcommand{\lepton}{\mathrm{Lept}}
\newcommand{\photon}{\mathrm{Ph}}
\newcommand{\lpath}{\mathrm{LPath}}
\newcommand{\vertex}{\mathrm{V}}
\newcommand{\edge}{\mathrm{E}}
\newcommand{\chains}{\mathrm{Chains}}
\newcommand{\ffch}{\mathrm{Ch}}

\newcommand{\csub}{C_{\mathrm{Sub}}}
\newcommand{\csubfseop}{C_{\mathrm{SubLSEOP}}}
\newcommand{\csubfseol}{C_{\mathrm{SubLSEOL}}}
\newcommand{\csubi}{C_{\mathrm{SubI}}}
\newcommand{\csubvop}{C_{\mathrm{SubVOP}}}
\newcommand{\csubvol}{C_{\mathrm{SubVOL}}}
\newcommand{\cbigf}{C_{\mathrm{BigF}}}
\newcommand{\cbigz}{C_{\mathrm{BigZ}}}
\newcommand{\cadd}{C_{\mathrm{Add}}}
\newcommand{\csat}{C_{\mathrm{Sat}}}

\begin{document}

\title{
Calculation of the total 10-th order QED contribution to the electron magnetic moment
}

\author{Sergey Volkov}%
 \email{sergey.volkov.1811@gmail.com, volkoff_sergey@mail.ru}
\affiliation{%
 Institute for Theoretical Physics, Karlsruhe Institute of Technology (KIT), \\
76128 Karlsruhe, Germany
}%


\begin{abstract}

The total 5-loop quantum electrodynamics universal contribution to the anomalous magnetic moments of the leptons was calculated by the author. The obtained value $A_1^{(10)}=5.891(61)$ provides the first complete verification of the previously known value obtained by T. Aoyama, M. Hayakawa, T. Kinoshita, M. Nio (AHKN). The discrepancy is $5 \sigma$. The computation includes the recalculation of the part that the author calculated in 2019 using a slightly different method and the calculation of the remaining part of the coefficient. 

A comparison with the AHKN values in 32 gauge-invariant classes is provided. In addition, the results are divided into 95 small gauge-invariant classes that subdivide the former ones. Such a detailzation is provided for the first time.

The method described in previous works of the author was used in general. However, the Monte Carlo integration method is new and is described in detail. 

Some useful technical information and information on numerical cancellations is also given.

\end{abstract}

\maketitle


\section{INTRODUCTION}

The electron anomalous magnetic moment $a_e$ is known with very high accuracy. The measurement \cite{experiment_electron_2022} provided the result
\begin{equation}\label{eq_experiment}
a_e=0.00115965218059(13)
\end{equation}

Standard Model predictions for $a_e$ use the following expression:
$$
a_e=a_e(\text{QED})+a_e(\text{hadronic})+a_e(\text{electroweak}),
$$
$$
a_e(\text{QED})=\sum_{n\geq 1} \left(\frac{\alpha}{\pi}\right)^n
a_e^{2n},
$$
$$
a_e^{2n}=A_1^{(2n)}+A_2^{(2n)}(m_e/m_{\mu})+A_2^{(2n)}(m_e/m_{\tau})+A_3^{(2n)}(m_e/m_{\mu},m_e/m_{\tau}),
$$
where $m_e,m_{\mu},m_{\tau}$ are the masses of the electron, the muon and the tau-lepton respectively, $\alpha$ is the fine-structure constant.

The universal QED terms $A_1^{(2n)}(\alpha/\pi)^n$ form the main contribution to $a_e$. The value
$$
A_1^{(2)}=0.5
$$
was determined by J. Schwinger in 1948 \cite{schwinger1,schwinger2}. The 2-loop contribution $A_1^{(4)}$ was mainly calculated by R. Karplus and N. Kroll \cite{karpluskroll}. However, this calculation had an error; the correct value
$$
A_1^{(4)}=-0.328478965579\ldots
$$
was presented independently in 1957 by A. Petermann \cite{analyt2_p} and C. Sommerfield \cite{analyt2_z}. The value of $A_1^{(6)}$ was being calculated in the 1970s by several research groups using numerical integration (\cite{carrollyao,carroll}; \cite{levinewright}; \cite{kinoshita_6}); the most accurate value $A_1^{(6)}=1.195\pm 0.026$ for that era was determined in 1974 by T. Kinoshita and P. Cvitanovi\'{c}; the uncertainty was caused by the statistical error of the Monte Carlo integration. At the same time, an analytical calculation of $A_1^{(6)}$ using computers was started. The final value 
$$
A_1^{(6)}=1.181241456\ldots
$$
was presented in 1996 by S. Laporta and E. Remiddi \cite{analyt3}. This value was the result of the efforts of many researchers (e.g., 
\cite{analyt_mi, analyt_b3, analyt_j, analyt_e, analyt_d, laporta_1993}
). The first numerical estimates for $A_1^{(8)}$ were given in 1981 by T. Kinoshita and W. B. Lindquist \cite{kinoshita_8_first}. The most precise value presented by T. Kinoshita's team, $A_1^{(8)}=-1.91298(84)$, was published in 2015 \cite{kinoshita_8_last}. This value was obtained by Monte Carlo integration. The semianalytic result of S. Laporta 
$$
A_1^{(8)}=-1.9122457649\ldots
$$
was presented in 2017 \cite{laporta_8}. These two calculations of $A_1^{(8)}$ agree well, as do another independent calculations ~\cite{smirnov_amm,rappl,volkov_method_details_2023}. First overall computation results for $A_1^{(10)}$ were published in 2012 by T. Aoyama, M. Hayakawa, T. Kinoshita, M. Nio (AHKN) in \cite{kinoshita_10_first}. The last value presented by these researchers in 2019 \cite{kinoshita_atoms} is
\begin{equation}\label{eq_kinoshita_10}
A_1^{(10)}[\text{AHKN}]=6.737(159).
\end{equation}
Up to this moment, the coefficient had not been verified and a significant computational error could be noticeable in experiments. In 2019, we recalculated the total contribution of the graphs without lepton loops to $A_1^{(10)}$ \cite{volkov_prd}; the value disagreed with the AHKN one.

In this paper we present new calculation results for $A_1^{(10)}$. Our value for the remaining part is
$$
A_1^{(10)}[\text{with lepton loops,Volkov}]=-0.9377(35).
$$
It agrees with the value
$$
A_1^{(10)}[\text{with lepton loops,AHKN}]=-0.933(17)
$$
from \cite{kinoshita_10_first}.

This value was obtained using a method slightly different from the one used in 2019 (see Sections \ref{sec_subtraction} and \ref{sec_monte_carlo}); the programs for the integrand code generation were also different. For this reason, we recalculated the contribution of the graphs without lepton loops with the aim of resolving the discrepancy. Our new value 
$$
A_1^{(10)}[\text{no lepton loops,Volkov,2024 only}]=6.857(81)
$$
agrees with our number published in \cite{volkov_5loops_2019}, but does not agree with that of the AHKN. The combined value is
$$
A_1^{(10)}[\text{no lepton loops,Volkov}]=6.828(60).
$$
Our total value
\begin{equation}\label{eq_volkov_10}
A_1^{(10)}[\text{Volkov}]=5.891(61)
\end{equation}
comes from our new calculation in combination with the old one. The discrepancy with (\ref{eq_kinoshita_10}) is $5 \sigma$.

At the moment the discrepancy in $A_1^{(10)}[\text{no lepton loops}]$ is unresolved, but independent calculations are coming \cite{kitanotakaura}. The values (\ref{eq_kinoshita_10}) and (\ref{eq_volkov_10}) in combination with the experimental value (\ref{eq_experiment}) and another known contributions \cite{kinoshita_atoms} lead to
\begin{equation}\label{eq_alpha_kinoshita}
\alpha^{-1}[a_e,\text{AHKN}]=137.0359991663(155)
\end{equation}
and
\begin{equation}\label{eq_alpha_volkov}
\alpha^{-1}[a_e,\text{Volkov}]=137.0359991595(155),
\end{equation}
respectively. The values obtained from the measured ratios of the atomic masses and the Planck constant
$$
\alpha^{-1}[\text{Rb-2011}]=137.035998996(85),
$$
$$
\alpha^{-1}[\text{Cs-2018}]=137.035999046(27),
$$
$$
\alpha^{-1}[\text{Rb-2020}]=137.035999206(11)
$$
come from \cite{alpha_rubidium}, \cite{alpha_cesium}, and \cite{alpha_rubidium_2020}, respectively. Note that $\alpha^{-1}[\text{Rb-2020}]$ is the largest among these three values and has a discrepancy of $5.4\sigma$ relative to $\alpha^{-1}[\text{Cs-2018}]$. The tensions with (\ref{eq_alpha_kinoshita}) are $1.97\sigma$, $3.86\sigma$, $2.09\sigma$; the corresponding tensions with (\ref{eq_alpha_volkov}) are $1.89\sigma$, $3.65 \sigma$, $2.45 \sigma$, respectively.

Other results exist for small classes of 5-loop and higher order graphs \cite{lautrup_bubble,ladders_12,muon_10_baikov}; they are in good agreement with the ones mentioned above.

The values for the individual classes and their comparison with the AHKN values as well as the hardware used are described in Section \ref{sec_results_detail}. The method of reduction to finite integrals used was described in detail in \cite{volkov_method_details_2023} and briefly recapitulated in Section \ref{sec_subtraction}; the chosen modification of the procedure was also described in this section. The integrals were evaluated numerically using Monte Carlo integration. We used a hand-made Monte Carlo. The general ideas of our Monte Carlo were described in \cite{volkov_prd}. It is based on predefined probability density functions obtained from the combinatorics of Feynman graphs. Since the procedure was developed in the previous works only for graphs without lepton loops, a serious modification and improvement is needed. This is described in Section \ref{sec_monte_carlo}. The information that can be used for the resolution of the discrepancy, another calculations, and theoretical investigations, is presented in Section \ref{sec_technical}.

\section{RESULTS IN DETAIL}\label{sec_results_detail}

We extract $a_e$ from QED Feynman graphs with $N_l=2$, $N_{\gamma}=1$, where we denote by $N_l$ and $N_{\gamma}$ the number of external lepton and photon lines in the graph; each graph may contain electron, muon, tau-lepton lines. We also assume that all graphs are one-particle irreducible and have no odd lepton loops (Furry's theorem). Since the contribution of a QED Feynman graph does not depend on the arrow directions, we use \emph{undirected} graphs everywhere in the calculations. By the number of Feynman graphs we always mean the number of undirected graphs.

We say that a vertex $v$ is \emph{incident} to a line $l$, if $v$ is one of the ends of $l$.

\begin{figure}[H]
	\begin{center}
		\includegraphics[width=180mm]{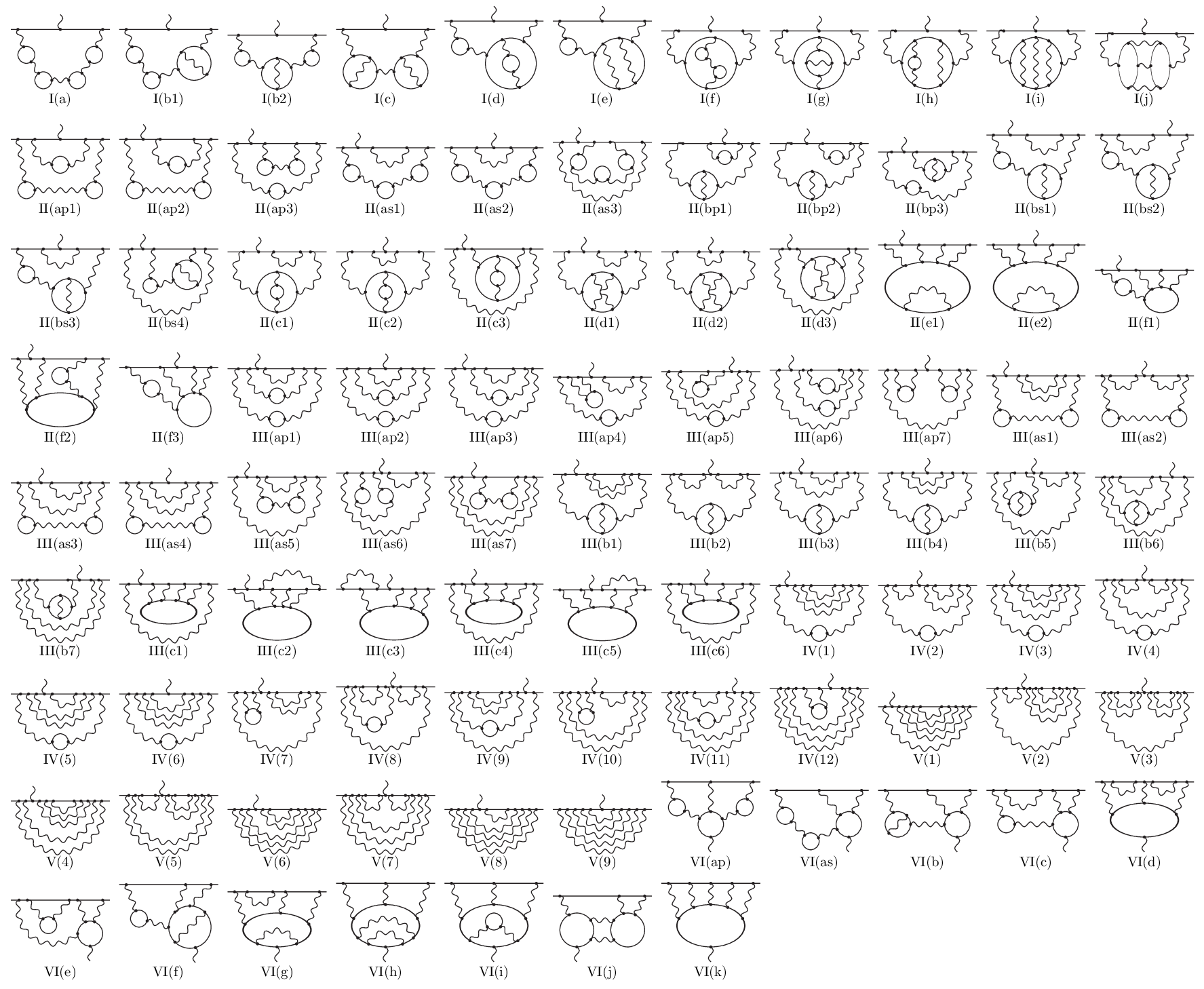}
		\caption{\label{fig_a1_5loops}Gauge-invariant classes contributing to $A_1^{(10)}$.}
	\end{center}
\end{figure}

Our contributions to $A_1^{(10)}$ are split into 95 gauge-invariant classes depicted in FIG. \ref{fig_a1_5loops}. One class is the set of all Feynman graphs that can be obtained from one picture by moving internal photon lines along lepton paths and loops, but without jumping over the vertex incident to the external photon line. The individual class contributions has to do with the in-place on-shell renormalization; the exact definition is given in \cite{volkov_method_details_2023}. The proof of their gauge invariance is given in \cite{cvitanovic_gauge}\footnote{The proof was only given for graphs without lepton loops. It looks the whole proof has not been published, but the fact is widely used; we believe that the ideas of \cite{cvitanovic_gauge} can be extended to the general case.}.

The newly calculated contributions to $A_1^{(10)}[\text{with lepton loops}]$ with the corresponding technical information can be found in Table \ref{table_calculations}. The calculation was performed on the supercomputer ``HoreKa'' (Karlsruhe, Germany). The GPUs NVidia A100 were used for the Monte Carlo integration (most of the time is taken by the integrand evaluation); The CPUs Intel Xeon Platinum 8368 were used for the control. To avoid biases caused by the pseuderandom number generators, two generators from the NVidia library were used: \texttt{MRG32k3a} and \texttt{Philox\char`_4x32\char`_10}. The separate results as well as the statistical averages are listed in the table. The number of the undirected graphs, of the Monte Carlo samples, the calculation time are presented in the last three columns. The corresponding contributions to $A_1^{(10)}[\text{no lepton loops}]$ are given in Table \ref{table_calculations_setv}. A comparison with the old values from \cite{volkov_5loops_2019} as well as the statistical mean values are presented in Table \ref{table_comparison_with_old_setv}. The new Monte Carlo algorithm described in Section \ref{sec_monte_carlo} takes approximately by $1.6$ times fewer samples to achieve the needed accuracy than the previous one \cite{volkov_5loops_2019}. However, this does not affect the computation time as much, since more samples require increased precisions in this case; see Section \ref{sec_technical}.

\begin{longtable}{ccccccc}\caption{The contributions of all gauge-invariant classes containing lepton loops to $A_1^{(10)}$, calculated with different random number generators (continued)} \\
\hline \hline Class & \texttt{Calc\char`_MRG} & \texttt{Calc\char`_Philox} & Average & $N_{\text{graph}}$ & $N_{\text{samples}}$ & Time (GPU-hours) \\ \hline  \endhead
\caption{The contributions of all gauge-invariant classes containing lepton loops to $A_1^{(10)}$, calculated with different random number generators; by $N_{\text{graph}}$ we denote the number of \emph{undirected} Feynman graphs}\label{table_calculations} \\
\hline \hline Class & \texttt{Calc\char`_MRG} & \texttt{Calc\char`_Philox} & Average & $N_{\text{graph}}$ & $N_{\text{samples}}$ & Time (GPU-hours) \\ \hline  \endfirsthead
\hline \endfoot  \hline \hline \endlastfoot
I(a) & 0.00047108(23) & 0.00047102(23) & 0.00047105(16) & 1 & $15\times 10^{11}$ & 30.9 \\
I(b1) & 0.0046717(16) & 0.0046715(16) & 0.0046715(11) & 2 & $12\times 10^{11}$ & 25.8 \\
I(b2) & 0.00233645(75) & 0.00233679(78) & 0.00233656(54) & 2 & $12\times 10^{11}$ & 26.3 \\
I(c) & 0.0234651(35) & 0.0234645(41) & 0.0234643(26) & 3 & $81\times 10^{10}$ & 22.8 \\
I(d) & 0.00380375(84) & 0.00380366(91) & 0.00380370(61) & 2 & $10^{12}$ & 22.8 \\
I(e) & 0.010293(16) & 0.010282(17) & 0.010289(11) & 9 & $29\times 10^{11}$ & 70.5 \\
I(f) & 0.00757078(67) & 0.00757131(66) & 0.00757106(47) & 2 & $84\times 10^{10}$ & 17.6 \\
I(g) & 0.0285715(26) & 0.0285678(24) & 0.0285696(18) & 4 & $15\times 10^{11}$ & 39.8 \\
I(h) & 0.001682(10) & 0.0016807(91) & 0.0016826(63) & 11 & $50\times 10^{11}$ & 114.5 \\
I(i) & 0.01727(43) & 0.01718(41) & 0.01726(29) & 39 & $85\times 10^{12}$ & 1416.2 \\
I(j) & 0.000401(11) & 0.0004044(79) & 0.0004038(63) & 2 & $74\times 10^{10}$ & 50.0 \\
II(ap1) & 0.0092489(11) & 0.0092448(14) & 0.00924758(84) & 3 & $79\times 10^{10}$ & 23.9 \\
II(ap2) & -0.0297767(17) & -0.0297768(16) & -0.0297767(11) & 2 & $85\times 10^{10}$ & 23.2 \\
II(ap3) & -0.0262168(17) & -0.0262195(14) & -0.0262183(11) & 2 & $49\times 10^{10}$ & 21.0 \\
II(as1) & -0.0466165(58) & -0.0466065(61) & -0.0466111(41) & 2 & $57\times 10^{10}$ & 19.3 \\
II(as2) & 0.0137932(21) & 0.0137901(23) & 0.0137916(15) & 3 & $13\times 10^{11}$ & 25.9 \\
II(as3) & -0.0299263(27) & -0.0299284(24) & -0.0299276(18) & 2 & $87\times 10^{10}$ & 13.4 \\
II(bp1) & 0.0326220(40) & 0.0326252(60) & 0.0326223(31) & 6 & $86\times 10^{10}$ & 28.7 \\
II(bp2) & -0.080010(10) & -0.0800158(91) & -0.0800129(67) & 4 & $56\times 10^{10}$ & 16.3 \\
II(bp3) & -0.083589(11) & -0.083617(13) & -0.0836017(85) & 4 & $29\times 10^{10}$ & 9.6 \\
II(bs1) & -0.125046(16) & -0.125060(15) & -0.125057(11) & 4 & $80\times 10^{10}$ & 34.0 \\
II(bs2) & -0.125034(21) & -0.125080(19) & -0.125056(14) & 4 & $67\times 10^{10}$ & 16.9 \\
II(bs3) & 0.088110(14) & 0.088118(19) & 0.088110(10) & 8 & $20\times 10^{11}$ & 40.4 \\
II(bs4) & -0.180651(20) & -0.180611(19) & -0.180629(14) & 8 & $13\times 10^{11}$ & 26.2 \\
II(c1) & -0.086935(12) & -0.086943(14) & -0.0869389(86) & 4 & $63\times 10^{10}$ & 16.2 \\
II(c2) & 0.0375586(58) & 0.0375556(85) & 0.0375573(46) & 6 & $10^{12}$ & 24.8 \\
II(c3) & -0.067120(11) & -0.067129(11) & -0.0671244(72) & 4 & $10^{12}$ & 20.9 \\
II(d1) & -0.18984(16) & -0.18988(16) & -0.18988(11) & 18 & $47\times 10^{11}$ & 110.9 \\
II(d2) & 0.11133(11) & 0.11137(10) & 0.111341(75) & 25 & $50\times 10^{11}$ & 100.1 \\
II(d3) & -0.16439(12) & -0.16438(12) & -0.164378(83) & 18 & $43\times 10^{11}$ & 76.1 \\
II(e1) & -1.13784(67) & -1.13735(64) & -1.13757(46) & 30 & $20\times 10^{11}$ & 730.1 \\
II(e2) & -0.20445(41) & -0.20506(39) & -0.20478(28) & 20 & $69\times 10^{10}$ & 306.6 \\
II(f1) & -0.23991(13) & -0.23990(12) & -0.239896(88) & 3 & $13\times 10^{10}$ & 43.9 \\
II(f2) & -1.91548(33) & -1.91474(32) & -1.91510(23) & 9 & $54\times 10^{10}$ & 186.6 \\
II(f3) & -0.28090(25) & -0.28019(24) & -0.28054(18) & 6 & $17\times 10^{10}$ & 118.1 \\
III(ap1) & 0.056150(65) & 0.056165(57) & 0.056161(42) & 10 & $30\times 10^{10}$ & 30.6 \\
III(ap2) & 0.251177(35) & 0.251187(41) & 0.251184(26) & 12 & $30\times 10^{10}$ & 39.5 \\
III(ap3) & 0.227933(70) & 0.227875(67) & 0.227894(47) & 20 & $82\times 10^{10}$ & 45.2 \\
III(ap4) & 0.143148(25) & 0.143129(25) & 0.143139(17) & 4 & $28\times 10^{10}$ & 26.4 \\
III(ap5) & 0.071158(55) & 0.071180(51) & 0.071170(37) & 20 & $69\times 10^{10}$ & 39.0 \\
III(ap6) & 0.164803(28) & 0.164778(32) & 0.164795(20) & 10 & $92\times 10^{10}$ & 28.8 \\
III(ap7) & 0.0275440(43) & 0.0275462(35) & 0.0275451(27) & 3 & $64\times 10^{10}$ & 34.3 \\
III(as1) & 0.054626(81) & 0.054572(76) & 0.054599(55) & 10 & $96\times 10^{10}$ & 31.3 \\
III(as2) & 0.225735(25) & 0.225727(25) & 0.225730(17) & 3 & $38\times 10^{10}$ & 24.4 \\
III(as3) & 0.05552(10) & 0.055647(92) & 0.055588(66) & 20 & $12\times 10^{11}$ & 52.9 \\
III(as4) & 0.289459(50) & 0.289388(50) & 0.289427(35) & 12 & $60\times 10^{10}$ & 26.1 \\
III(as5) & 0.439958(81) & 0.440098(78) & 0.440024(55) & 20 & $16\times 10^{11}$ & 48.2 \\
III(as6) & 0.102922(30) & 0.103027(28) & 0.102979(21) & 4 & $93\times 10^{10}$ & 26.6 \\
III(as7) & 0.017024(52) & 0.017036(49) & 0.017022(35) & 10 & $81\times 10^{10}$ & 22.2 \\
III(b1) & 0.35844(19) & 0.35866(18) & 0.35850(13) & 20 & $22\times 10^{11}$ & 104.5 \\
III(b2) & 0.55074(11) & 0.55070(11) & 0.550730(74) & 6 & $51\times 10^{10}$ & 28.9 \\
III(b3) & -0.21853(25) & -0.21798(23) & -0.21825(16) & 40 & $36\times 10^{11}$ & 164.2 \\
III(b4) & 0.91605(13) & 0.91600(13) & 0.916024(91) & 24 & $17\times 10^{11}$ & 69.0 \\
III(b5) & 0.43822(15) & 0.43817(15) & 0.43819(10) & 8 & $13\times 10^{11}$ & 48.3 \\
III(b6) & 1.35328(23) & 1.35317(22) & 1.35323(15) & 40 & $42\times 10^{11}$ & 146.5 \\
III(b7) & -0.07136(15) & -0.07093(14) & -0.07113(10) & 20 & $21\times 10^{11}$ & 65.4 \\
III(c1) & 1.86016(70) & 1.86020(67) & 1.86018(48) & 12 & $20\times 10^{11}$ & 749.5 \\
III(c2) & 3.2810(13) & 3.2801(13) & 3.28050(91) & 27 & $36\times 10^{11}$ & 2590.2 \\
III(c3) & 0.27186(70) & 0.27149(66) & 0.27166(48) & 6 & $12\times 10^{11}$ & 705.1 \\
III(c4) & -0.7592(11) & -0.7596(11) & -0.75943(79) & 24 & $23\times 10^{11}$ & 1982.5 \\
III(c5) & 0.22273(75) & 0.22339(71) & 0.22308(51) & 15 & $13\times 10^{11}$ & 858.0 \\
III(c6) & 0.04462(66) & 0.04334(64) & 0.04392(46) & 18 & $10^{12}$ & 744.0 \\
IV(1) & -0.51577(63) & -0.51599(59) & -0.51586(43) & 74 & $53\times 10^{11}$ & 734.9 \\
IV(2) & -0.64897(40) & -0.64858(38) & -0.64878(28) & 20 & $18\times 10^{11}$ & 276.2 \\
IV(3) & -1.14814(88) & -1.14833(84) & -1.14824(60) & 148 & $78\times 10^{11}$ & 1544.6 \\
IV(4) & 1.19585(56) & 1.19608(53) & 1.19593(38) & 55 & $32\times 10^{11}$ & 594.6 \\
IV(5) & -1.52733(88) & -1.52833(83) & -1.52785(60) & 162 & $74\times 10^{11}$ & 1605.4 \\
IV(6) & 0.50510(44) & 0.50550(42) & 0.50531(30) & 56 & $19\times 10^{11}$ & 444.5 \\
IV(7) & -0.19297(38) & -0.19292(36) & -0.19295(26) & 20 & $26\times 10^{11}$ & 254.9 \\
IV(8) & -0.78534(55) & -0.78358(53) & -0.78444(38) & 40 & $61\times 10^{11}$ & 526.5 \\
IV(9) & -4.4900(11) & -4.4904(10) & -4.49029(73) & 222 & $21\times 10^{12}$ & 2075.6 \\
IV(10) & 0.19705(47) & 0.19693(44) & 0.19698(32) & 50 & $40\times 10^{11}$ & 432.6 \\
IV(11) & 0.05155(81) & 0.05307(76) & 0.05233(55) & 148 & $12\times 10^{12}$ & 1317.3 \\
IV(12) & -0.37235(48) & -0.37267(44) & -0.37250(31) & 54 & $35\times 10^{11}$ & 476.4 \\
VI(ap) & 0.482912(86) & 0.483009(83) & 0.482955(59) & 5 & $14\times 10^{10}$ & 41.4 \\
VI(as) & 0.558642(83) & 0.558526(80) & 0.558582(57) & 5 & $27\times 10^{10}$ & 37.4 \\
VI(b) & 1.34680(17) & 1.34710(16) & 1.34697(11) & 10 & $10^{12}$ & 64.7 \\
VI(c) & -2.53393(71) & -2.53241(67) & -2.53312(49) & 37 & $30\times 10^{11}$ & 839.9 \\
VI(d) & 1.8426(32) & 1.8506(30) & 1.8468(22) & 127 & $41\times 10^{12}$ & 14465.2 \\
VI(e) & -0.43170(25) & -0.43094(24) & -0.43129(17) & 13 & $55\times 10^{10}$ & 125.5 \\
VI(f) & 0.77156(34) & 0.77158(32) & 0.77154(23) & 46 & $12\times 10^{11}$ & 295.7 \\
VI(g) & -1.5962(14) & -1.5968(14) & -1.5965(10) & 122 & $10^{13}$ & 3221.2 \\
VI(h) & 0.1855(10) & 0.18576(93) & 0.18554(68) & 162 & $12\times 10^{12}$ & 1913.8 \\
VI(i) & -0.04398(15) & -0.04392(14) & -0.04396(10) & 16 & $62\times 10^{10}$ & 68.3 \\
VI(j) & -0.22921(63) & -0.22917(60) & -0.22920(43) & 9 & $63\times 10^{10}$ & 644.1 \\
VI(k) & 0.67989(58) & 0.67970(56) & 0.67974(39) & 32 & $32\times 10^{11}$ & 611.5 \\
TOTAL & -0.9438(51) & -0.9314(49) & -0.9377(35) & 2323 & $32\times 10^{13}$ & 45214.1 \\
\end{longtable}

\begin{longtable}{ccccccc}\caption{The contributions of all gauge-invariant classes without lepton loops to $A_1^{(10)}$, calculated with different random number generators (continued)} \\
\hline \hline Class & \texttt{Calc\char`_MRG} & \texttt{Calc\char`_Philox} & Average & $N_{\text{graph}}$ & $N_{\text{samples}}$ & Time (GPU-hours) \\ \hline  \endhead
\caption{The contributions of all gauge-invariant classes without lepton loops to $A_1^{(10)}$, calculated with different random number generators; by $N_{\text{graph}}$ we denote the number of \emph{undirected} Feynman graphs}\label{table_calculations_setv} \\
\hline \hline Class & \texttt{Calc\char`_MRG\char`_V} & \texttt{Calc\char`_Philox\char`_V} & Average & $N_{\text{graph}}$ & $N_{\text{samples}}$ & Time (GPU-hours) \\ \hline  \endfirsthead
\hline \endfoot  \hline \hline \endlastfoot
V(1) & 6.175(30) & 6.167(30) & 6.169(20) & 706 & $28\times 10^{12}$ & 4428.0 \\
V(2) & 0.962(44) & 0.966(45) & 0.964(31) & 148 & $40\times 10^{12}$ & 5436.2 \\
V(3) & 0.342(29) & 0.326(29) & 0.334(20) & 55 & $13\times 10^{12}$ & 1942.6 \\
V(4) & -0.832(53) & -0.748(53) & -0.789(37) & 706 & $54\times 10^{12}$ & 9497.4 \\
V(5) & -2.139(49) & -2.174(49) & -2.157(34) & 370 & $42\times 10^{12}$ & 8028.9 \\
V(6) & -0.402(51) & -0.426(51) & -0.415(35) & 558 & $32\times 10^{12}$ & 9676.1 \\
V(7) & 2.677(33) & 2.612(33) & 2.644(23) & 261 & $18\times 10^{12}$ & 4360.5 \\
V(8) & -1.005(31) & -0.942(31) & -0.975(21) & 336 & $14\times 10^{12}$ & 4300.8 \\
V(9) & 1.0807(92) & 1.0795(92) & 1.0807(63) & 73 & $21\times 10^{11}$ & 560.8 \\
TOTAL & 6.86(12) & 6.86(12) & 6.857(81) & 3213 & $24\times 10^{13}$ & 48231.4 \\
\end{longtable}

\begin{longtable}{cccc}\caption{Comparison of the contributions of the classes without lepton loops to $A_1^{(10)}$ with the values from \cite{volkov_5loops_2019}, statistical averages (continued)} \\
\hline \hline Class & New value & Old value (2019) & Average \\ \hline  \endhead
\caption{Comparison of the contributions of the classes without lepton loops to $A_1^{(10)}$ with the values from \cite{volkov_5loops_2019}, statistical averages}\label{table_comparison_with_old_setv} \\
\hline \hline Class & New value & Old value (2019) & Average \\ \hline  \endfirsthead
\hline \endfoot  \hline \hline \endlastfoot
V(1) & 6.169(20) & 6.157(33) & 6.166(17) \\
V(2) & 0.964(31) & 0.970(33) & 0.967(23) \\
V(3) & 0.334(20) & 0.315(20) & 0.325(15) \\
V(4) & -0.789(37) & -0.754(42) & -0.774(28) \\
V(5) & -2.157(34) & -2.165(36) & -2.161(25) \\
V(6) & -0.415(35) & -0.403(34) & -0.409(24) \\
V(7) & 2.644(23) & 2.625(24) & 2.635(17) \\
V(8) & -0.975(21) & -1.011(21) & -0.993(15) \\
V(9) & 1.0807(63) & 1.0902(62) & 1.0855(44) \\
TOTAL & 6.857(81) & 6.793(90) & 6.828(60)
\end{longtable}

The comparison of the averaged values with the AHKN values \cite{kinoshita_10_first, kinoshita_atoms} is given in Table \ref{table_comparison_ahkn}. Each class is the union of the classes from FIG. \ref{fig_a1_5loops} with the corresponding names; the value is the sum of the corresponding numbers. For example, II(b) consists of II(bp1), II(bp2), II(bp3), II(bs1), II(bs2), II(bs3), II(bs4); here ``p'' means ``parallel'', ``s'' means ``sequential'' (it concerns the placement of the photon self-energy insertions). The names are taken from \cite{kinoshita_10_first}. The results agree very well in all sets except Set V. In the latter, a discrepancy of $5 \sigma$ remains.

\begin{longtable}{ccc}\caption{Comparison of the contributions to $A_1^{(10)}$ with the AHKN values \cite{kinoshita_10_first, kinoshita_atoms} (continued)} \\
\hline \hline Class & This value & AHKN value \\ \hline  \endhead
\caption{Comparison of the contributions to $A_1^{(10)}$ with the AHKN values \cite{kinoshita_10_first, kinoshita_atoms}}\label{table_comparison_ahkn} \\
\hline \hline Class & This value & AHKN value \\ \hline  \endfirsthead
\hline \endfoot  \hline \hline \endlastfoot
I(a) & 0.00047105(16) & 0.000470940(60) \\
I(b) & 0.0070081(12) & 0.00701080(70) \\
I(c) & 0.0234643(26) & 0.0234680(20) \\
I(d) & 0.00380370(61) & 0.00380170(50) \\
I(e) & 0.010289(11) & 0.0102960(40) \\
I(f) & 0.00757106(47) & 0.0075684(20) \\
I(g) & 0.0285696(18) & 0.0285690(60) \\
I(h) & 0.0016826(63) & 0.001696(13) \\
I(i) & 0.01726(29) & 0.01747(11) \\
I(j) & 0.0004038(63) & 0.0003975(18) \\
II(a) & -0.1094945(50) & -0.109495(23) \\
II(b) & -0.473625(27) & -0.473559(84) \\
II(c) & -0.116506(12) & -0.116489(32) \\
II(d) & -0.24291(15) & -0.24300(29) \\
II(e) & -1.34235(54) & -1.3449(10) \\
II(f) & -2.43553(30) & -2.4336(15) \\
III(a) & 2.12726(14) & 2.12733(17) \\
III(b) & 3.32730(32) & 3.32712(45) \\
III(c) & 4.9199(15) & 4.921(11) \\
IV & -7.7303(16) & -7.7296(48) \\
V & 6.828(60) & 7.670(159) \\
VI(a) & 1.041537(82) & 1.04132(19) \\
VI(b) & 1.34697(11) & 1.34699(28) \\
VI(c) & -2.53312(49) & -2.5289(28) \\
VI(d) & 1.8468(22) & 1.8467(70) \\
VI(e) & -0.43129(17) & -0.43120(70) \\
VI(f) & 0.77154(23) & 0.7703(22) \\
VI(g) & -1.5965(10) & -1.5904(63) \\
VI(h) & 0.18554(68) & 0.1792(39) \\
VI(i) & -0.04396(10) & -0.0438(12) \\
VI(j) & -0.22920(43) & -0.2288(18) \\
VI(k) & 0.67974(39) & 0.6802(38) \\
\end{longtable}

\section{REDUCTION TO FINITE INTEGRALS}\label{sec_subtraction}

We work in the unit system with $\hbar=c=1$, the factors of $4\pi$ appear in the fine-structure constant: $\alpha=e^2/(4\pi)$, the tensor $g_{\mu\nu}$ corresponds to the signature $(+,-,-,-)$, the Dirac matrices fulfill the condition $\gamma_{\mu}\gamma_{\nu}+\gamma_{\nu}\gamma_{\mu}=2g_{\mu\nu}$. 

We work in the Feynman gauge with the propagators 
$$
\frac{i(\slashed{q}+m)}{q^2-m^2+i\varepsilon},\quad \frac{-ig_{\mu\nu}}{q^2+i\varepsilon}
$$
for leptons and photons, where $m$ is the lepton mass.

The method of reduction of $A_1^{(2n)}$ to finite integrals is described in detail in \cite{volkov_method_details_2023}. It gives one Feynman parametric integral for each Feynman graph. The final value is simply the sum of these integrals. To make each integral finite, a subtraction procedure with linear operators applied to the Feynman amplitudes of ultraviolet (UV) divergent subgraphs is utilized before arriving at the Feynman parameters. The use of direct subtraction of divergences under the integral sign is very important at the 5-loop level: the amount of  computer resources required would be otherwise astronomically large. We focus here only on the aspects that are important for the current implementation.

There are the following types of UV-divergent subgraphs\footnote{We consider only those subgraphs which are one-particle irreducible and contain all lines connecting the vertexes of the given subgraph; since odd lepton loops are forbidden, a UV-divergent subgraph is one-particle irreducible if and only if it is amputated.} in QED Feynman graphs: \emph{lepton self-energy} subgraphs ($N_l=2$, $N_{\gamma}=0$), \emph{vertexlike} subgraphs ($N_l=2$, $N_{\gamma}=1$), \emph{photon self-energy} subgraphs ($N_l=0$, $N_{\gamma}=2$), \emph{photon-photon scattering} subgraphs\footnote{Photon-photon scattering subgraph divergences cancel out in the final result without subtraction, but they remain in the individual graphs.} ($N_l=0$, $N_{\gamma}=4$).

We use standard operators like the anomalous magnetic moment projector and those that define the on-shell renormalization, but also special operators $U$ that are applied to the Feynman amplitudes of vertexlike and lepton self-energy graphs. In terms of \cite{volkov_method_details_2023}, we put $U_1=U_2=U_3=U$. 

The Feynman graphs contributing to the anomalous magnetic moment contain ultraviolet and infrared (IR) divergences (and the mixed ones). All divergences cancel out in the final result since the on-shell renormalization is applied. However, any direct BPHZ-like implementation of the on-shell renormalization leaves individual graphs IR-divergent. To remove the IR divergences, the combinatorics of the subtraction must be modified (it is explained in detail in \cite{volkov_method_details_2023}). The operators $U$ are used to remove UV divergences without the generation of additional IR divergences (and sometimes to remove IR divergences without generating additional UV divergences). All contributions of the terms containing $U$-operators are cancelled in the final result. In reality, the $U$-operators only redistribute the divergences between different graphs and regions in Feynman parametric space, this is why we have a free hand in the definition of the $U$-operators.

If $\Gamma_{\mu}(p,q)$ is a vertexlike Feynman amplitude, $p-\frac{q}{2}$, $p+\frac{q}{2}$ are incoming and outgoing lepton momenta, $q$ is the photon momentum, 
$$
\Gamma_{\mu}(p,0)=a(p^2)\gamma_{\mu} + b(p^2)p_{\mu} + c(p^2) \slashed{p}p_{\mu} + d(p^2)(\slashed{p}\gamma_{\mu}-\gamma_{\mu}\slashed{p})
$$
is satisfied, then put by definition
\begin{equation}\label{eq_u_vertex}
(U\Gamma)_{\mu}(p,q)=a(M^2)\gamma_{\mu},
\end{equation}
where $M^2$ is an arbitrary number. Similarly, for a lepton self-energy Feynman amplitude $\Sigma(p)=r(p^2)+s(p^2)\slashed{p}$ we define
\begin{equation}\label{eq_u_self_energy}
(U\Sigma)(p)=r(m^2)+s(m^2)m+s(M^2)(\slashed{p}-m).
\end{equation}
The preservation of the Ward identity plays an important role in the equivalence of the subtraction method used with the on-shell renormalization: If $\Sigma$ and $\Gamma$ satisfy
$$
\Gamma_{\mu}(p,0)=-e\frac{\partial \Sigma(p)}{\partial p^{\mu}},
$$
where $e$ is the lepton charge, then
$$
(U\Gamma)_{\mu}(p,0)=-e\frac{\partial (U \Sigma)(p)}{\partial p^{\mu}}
$$
is also satisfied. In old calculations we used $M^2=m^2$. Here we use $M^2=-m^2$; this slightly reduces inter-graph cancellations; see Section \ref{sec_technical} and \cite{volkov_acat_2021}.

For photon self-energy Feynman amplitudes $\Pi_{\mu\nu}(p^2)=h_1(p^2)g_{\mu\nu}+h_2(p^2) p_{\mu} p_{\nu}$ we utilize the substitution
$$
\Pi_{\mu\nu}(p^2) \rightarrow p^2 g_{\mu\nu}[h_2(0)-h_2(p^2)]
$$
instead of subtractions. 
This trick makes the expressions shorter and has been widely used in literature; see, for example, \cite{kinoshita_infrared}.

The divergences associated with the photon-photon scattering subgraphs are removed by subtracting the value at zero momenta.

\section{MONTE CARLO INTEGRATION}\label{sec_monte_carlo}

\subsection{General idea}

Suppose we have a Feynman parametric integral
$$
\int_{z_1,\ldots,z_K>0} I(z_1,\ldots,z_K)\delta(z_1+\ldots+z_K-1) dz_1\ldots dz_K
$$
corresponding to a graph of $A_1^{(2n)}$. Here $K=3n-1$: we use a trick to reduce the number from $3n$ to $3n-1$ (see \cite{volkov_prd}); each $z_i$ corresponds to an internal graph line, except one corresponding to two lepton lines adjoining the external photon line.

For the integration we use the predefined probability density functions of the form
$$
g(\underline{z}) = C_{\mcmain} \times g_0(\underline{z}) + C_{\mcmin} \times g_{\mcmin}(\underline{z}) + C_{\mcuniform} \times g_{\mcuniform}(\underline{z}) + C_{\mcmodify} \times \frac{1}{N}\sum_{i=1}^N g_i(\underline{z}),
$$
where $\underline{z}=(z_1,\ldots,z_K)$; $C_{\mcmain}, C_{\mcmin}, C_{\mcuniform}, C_{\mcmodify}$ are arbitrary nonnegative numbers, 
$$
C_{\mcmain} + C_{\mcmin} + C_{\mcuniform} + C_{\mcmodify}=1;
$$
$N$ is an arbitrary natural number.

The function $g_0$ has the form
$$
g_0(\underline{z}) = C \times \frac{\prod_{l=2}^K \left( z_{j_l} / z_{j_{l-1}} \right)^{\ffdeg_0(\{j_l,j_{l+1},\ldots,j_K\})}}{z_1 z_2 \ldots z_K},
$$
where
\begin{itemize}
\item $(j_1,\ldots,j_K)$ is a permutation of $1,\ldots,K$ such that $z_{j_1}\geq z_{j_2}\geq\ldots\geq z_{j_K}$ (this splitting of the integration area is called the Hepp sectors);
\item $\ffdeg_0(s)$ are arbitrary positive real numbers defined for any set $s\subseteq \{1,2,\ldots,K\}$ (except the empty and full sets);
\item $C$ is a constant such that
$$
\int_{z_1,\ldots,z_K>0} g_0(\underline{z}) \delta(z_1+\ldots+z_K-1) dz_1\ldots dz_K = 1.
$$
\end{itemize}
The algorithm for determining $\ffdeg_0$ for each set of internal graph lines is described in Section \ref{subsec_mc_main}; each set $s\subseteq \{1,\ldots,K\}$ can easily be mapped to the set of internal graph lines.

The functions $g_i$, $1\leq i\leq N$, and $g_{\mcmin}$ have the same form as $g_0$, but with the numbers $\ffdeg_i(s)$ and $\ffdeg_{\mcmin}(s)$ instead of $\ffdeg_0(s)$ (and with different $C$). We use
$$
\ffdeg_i(s)=\max(\mcdegsat_i, \mcdegmul_i \ffdeg_0(s) - \mcdegsub_i),
$$
$$
\ffdeg_{\mcmin}(s)=D,
$$
where $D$, $\mcdegsat_i$, $\mcdegmul_i$, $\mcdegsub_i$ ($1\leq i\leq N$) are arbitrary positive real numbers. The function $g_{\mcuniform}$ corresponds to the uniform distribution (constant).

We use the following constant values:
$$
C_{\mcmin}=0.002,\quad C_{\mcuniform}=0.06,\quad C_{\mcmodify}=0.02,
$$
$$
C_{\mcmain}=1-C_{\mcmin}-C_{\mcuniform}-C_{\mcmodify}=0.918,
$$
$$
D=0.75,\quad N=2,
$$
$$
\mcdegsat_1=0.6,\quad \mcdegmul_1=1.0,\quad \mcdegsub_1=1.3,
$$
$$
\mcdegsat_2=1.0,\quad \mcdegmul_2=0.4,\quad \mcdegsub_2=0.6.
$$

The algorithm for generating the random samples with the probability density functions of this form is described in \cite{volkov_prd}. The function $g_0(\underline{z})$ is designed in such a way to be ``near'' to $C|I(\underline{z})|$. The remaining terms of $g(\underline{z})$ are stabilization terms; they prevent from significant underestimation of $|I(\underline{z})|$, which can lead to poor Monte Carlo convergence.

The choice of $\mcdegsat_i$, $\mcdegmul_i$, $\mcdegsub_i$ is the result of experiments after the discovery of poor convergence for some Feynman graphs from Set I(i).

We have several integrals in our calculation. They are evaluated simultaneously by blocks of samples in random order. The probabilities of choosing each integral are adjusted in real time to make the overall convergence as fast as possible. The algorithm is described in \cite{volkov_5loops_2019} (Section IV.A).

\subsection{Obtaining $\ffdeg_0(s)$}\label{subsec_mc_main}

\subsubsection{Preliminaries}

This construction is based on \cite{volkov_iclos_npb} (ideologically\footnote{In fact, a serious improvement of the theory is needed to justify this procedure.}) and on numerical experiments with 4-loop graphs. It would be great if the integral
$$
\int_{z_1,\ldots,z_K>0} \frac{|I(\underline{z})|^2}{g_0(\underline{z})}\delta(z_1+\ldots+z_K-1)dz_1\ldots dz_K
$$
was finite and not so large (see \cite{volkov_prd}). However, we can not guarantee this for all orders; moreover, examples of high order are known that result in an infinite integral.

To write a formula for $\ffdeg_0(s)$, we need some additional definitions.

Two subgraphs are called \emph{overlapping} if they are not contained in each other and the intersection of their line sets is not empty.

A set of subgraphs of a graph is called a \emph{forest} if any two elements of this set do not overlap.

For a vertexlike graph $G$, we denote by $\forests[G]$  the \emph{set} of all forests $F$ consisting of UV-divergent subgraphs of $G$ and satisfying the condition $G\in F$.

If $s$ is some set of internal graph lines, by $\nloops(s)$ we denote the number of \emph{independent loops} in $s$; by $\lepton(s)$ we denote the set of all \emph{lepton lines} in $s$; by $\photon(s)$ we denote the set of all \emph{photon lines} in $s$; by $\vertex(s)$ we denote the set of all \emph{vertices} incident to at least one line in $s$.

If $G$ is an arbitrary Feynman graph, we denote by $\edge[G]$ the set of all \emph{internal lines} of $G$, by $\vertex[G]$ we denote the set of all vertices of $G$. If additionally $G$ has two external leptons, by $\lpath[G]$ we denote the \emph{lepton path} connecting the external lepton lines of $G$ (as a set of lines).

A subgraph $G'$ of a graph $G$ is said to lie on $\lpath[G]$, if $G'$ has external lepton lines, and each of these lines is in $\lpath[G]$ or coincides with an external line of $G$.

A set $s\subseteq\edge[G]$ is called \emph{cyclo-complete}, if
$$
(\lepton(\edge[G])\backslash \lpath[G])\subseteq s.
$$

\subsubsection{Auxiliary functions}

For convenience, we mark each new definition of global scope with bold lettering.

\begin{figure}[H]
	\begin{center}
		\includegraphics[width=80mm]{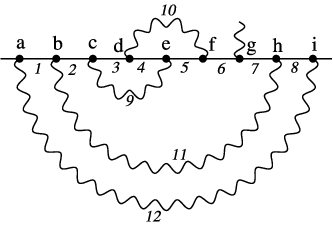}
		\caption{\label{fig_example_noloops}Example of a Feynman graph without lepton loops}
	\end{center}
\end{figure}

The \emph{ultraviolet degree of divergence} is \textbf{defined} as
$$
\omega_G(s) = 2 \times \nloops(s) - |s| + \frac{1}{2}|\lepton(s)|,
$$
where $s\subseteq \edge[G]$. For example, for the graph $G$ in FIG. \ref{fig_example_noloops} we have
$$
\omega_G(\{1,2,3,4,5,6,7,8,9,10,11\}) = 6-11+4=-1,
$$
$$
\omega_G(\{1,3,4,5,8,9,10,11,12\})= 6 -9+2.5 = -0.5.
$$

A graph $G''$ is \textbf{called} a \emph{child} of $G'$ in a forest $F$ ($G'\in F$), if $G''$ is a maximal (with respect to inclusion) element of $F$ properly contained in $G'$.

If $f\in\forests[G]$ and $G'\in F$, then by $G'/F$ we \textbf{denote} the graph that is obtained from $G'$ by shrinking all children of $G'$ in $F$ to points. We will also use the symbols like $\omega$ for graphs like $G'/F$ and for sets $s\subseteq\edge[G]$. This means that we really apply it to the set $s'$ that is the intersection of $s$ and the set of all internal lines of $G'/F$. 

Let us consider an example. Suppose $G$ is the graph from FIG. \ref{fig_example_noloops}, the UV-divergent subgraphs are
$$
G_1=\mathrm{bcdefgh},\ G_2=\mathrm{cdef},\ G_3=\mathrm{cde},\ G_4=\mathrm{def}
$$
(each subgraph is given by vertex enumeration). Suppose
$$
F=\{G,G_1,G_2,G_3\}.
$$
In this case,
$$
\omega_{G_1/F}(\{1,2,3,4,5,6,7,8,9,10,11\}) = \omega_{G_1/F}(\{2,6,7,11\})
$$
$$
 = 2 - 4 + 1.5 = -0.5
$$
($G_2$ is the only child of $G_1$ in $F$).

\textbf{Let us define} $\overline{\omega}_G(s)$, $s\subseteq \edge[G]$. It is defined only for $G$ with two external fermion lines (vertexlike or lepton self-energy graphs). The set $s\cap \lpath[G]$ can be considered as a union of nonintersecting paths. Suppose $I_1,\ldots,I_l$ are the sets of vertices corresponding to these paths (it is obvious that $|I_j|\geq 2$ for any $j$). The empty set gives $l=0$. By definition, put
$$
I_0=\lpath[G]\backslash(I_1\cup \ldots \cup I_l).
$$
We say that two vertices $v_1,v_2$ from $\vertex(\lpath[G])$ are 1-equivalent, if there is a path in $\edge[G]\backslash \lpath[G]$ from $v_1$ to $v_2$ (we consider all lines as undirected). Suppose $B_1,B_2,\ldots,B_r$ are classes of equivalence of $\vertex(\lpath[G])$ with respect to 1-equivalence. We say that $v_1$ and $v_2$ are 2-equivalent, if there is a path in $s \backslash \lpath[G]$ from $v_1$ to $v_2$ (we consider all lines as undirected). By $N_{jb}$ we denote the number of equivalence classes in $I_j\cap B_b$ with respect to 2-equivalence (the number is $0$ if the set is empty). By definition, put
$$
a_{jb}=\begin{cases} 
1, \text{ if } 0<N_{jb} < \sum_{i=0}^l N_{ib}, \\
\frac{1}{2}, \text{ if } 1=N_{jb}=\sum_{i=0}^l N_{ib}, \\
0,\text{ in the other cases}.
\end{cases}
$$
The function $\overline{\omega}_G$ is defined as
$$
\overline{\omega}_G(s) = 2\times \nloops(s) - |s| + \frac{1}{2}|\lepton(s) \backslash \lpath[G]| + \frac{1}{2} \sum_{j=1}^l |I_j| + \frac{1}{2}\sum_{j=1}^l \sum_{b=1}^r  N_{jb} - \sum_{j=1}^l \sum_{b=1}^r  a_{jb}  - \frac{1}{2}.
$$

\begin{figure}[H]
	\begin{center}
		\includegraphics[width=50mm]{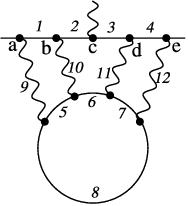}
		\caption{\label{fig_example_4ph}Example of a Feynman graph with a lepton loop of length 4}
	\end{center}
\end{figure}

Let us examine several examples. If $G$ is the graph from FIG. \ref{fig_example_4ph}, we have
$$
r=2,\quad B_1=\{\mathrm{a,b,d,e}\},\quad B_2=\{\mathrm{c}\}.
$$

For $s=\{1,2,3,9,10,11,12\}$, we have
$$
l=1,\quad I_1=\{\mathrm{a,b,c,d}\},\quad I_0=\{\mathrm{e}\},
$$
$$
I_0\cap B_1=\{\mathrm{e}\},\ N_{01}=1,\quad I_1\cap B_1=\{\mathrm{a,b,d}\},\ N_{11}=3,
$$
$$
I_0\cap B_2=\emptyset,\ N_{02}=0,\quad I_1\cap B_2=\{\mathrm{c}\},\ N_{12}=1,
$$
$$
a_{11}=1,\quad a_{12}=\frac{1}{2},
$$
$$
\overline{\omega}_G(s)=0-7+0+2+2-1.5-0.5=-5
$$
For $s=\{1,2,3,5,6,7,8,9,10,11,12\}$ we have the same $l,I_0,I_1$,
$$
N_{01}=1,\quad N_{11}=1,\quad N_{02}=0,\quad N_{12}=1,
$$
$a_{jb}$ are the same,
$$
\overline{\omega}_G(s)=6-11+2+2+1-1.5-0.5=-2
$$
For $s=\{1,2,3,6,9,10,11,12\}$ we have the same $l,I_0,I_1$, 
$$
N_{01}=1,\quad N_{11}=2,\quad N_{02}=0,\quad N_{12}=1,
$$
$a_{jb}$ are the same,
$$
\overline{\omega}_G(s)=2-8+0.5+2+1.5-1.5-0.5=-4.
$$
For $s=\{1,2,3,4,5,6,7,8\}$ we have
$$
l=1,\quad I_1=\{\mathrm{a,b,c,d,e}\},\quad I_0=\emptyset,
$$
$$
N_{01}=0,\quad N_{11}=4,\quad N_{02}=0,\quad N_{12}=1,
$$
$$
a_{11}=0,\quad a_{12}=\frac{1}{2},
$$
$$
\overline{\omega}_G(s)=2-8+2+2.5+2.5-0.5-0.5=0.
$$
For $s=\{1,2,3,4,5,6,7,8,9,10,11\}$ we have the same $l,I_0,I_1$,
$$
N_{01}=0,\quad N_{11}=2,\quad N_{02}=0,\quad N_{12}=1,
$$
$a_{jb}$ are the same,
$$
\overline{\omega}_G(s)=6-11+2+2.5+1.5-0.5-0.5=0.
$$
For $s=\{1,2,3,4,5,6,7,8,9,10,11,12\}$ we have the same $l,I_0,I_1$
$$
N_{01}=0,\quad N_{11}=1,\quad N_{02}=0,\quad N_{12}=1,
$$
$$
a_{11}=\frac{1}{2},\quad a_{12}=\frac{1}{2},
$$
$$
\overline{\omega}_G(s)=8-12+2+2.5+1-1-0.5=0.
$$

\begin{figure}[H]
	\begin{center}
		\includegraphics[width=30mm]{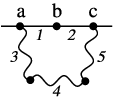}
		\caption{\label{fig_example_special}Example of a Feynman graph with special vertices}
	\end{center}
\end{figure}

If $G$ is the graph from FIG. \ref{fig_example_special} containing special vertices, we have
$$
r=2,\quad B_1=\{\mathrm{a,c}\},\quad B_2=\{\mathrm{b}\}.
$$

For $s=\emptyset$, we have
$$
l=0,\quad I_0=\{\mathrm{a,b,c}\},
$$
$$
\overline{\omega}_G(s) = 0-0+0+0+0-0-0.5 = -0.5.
$$
For $s=\{3,4,5\}$, we have the same $l=0$,
$$
\overline{\omega}_G(s) = 0-3+0+0+0-0-0.5 = -3.5.
$$
For $s=\{1,2\}$, we have
$$
l=1,\quad I_1=\{\mathrm{a,b,c}\},\quad I_0=\emptyset,
$$
$$
N_{01}=0,\quad N_{11}=2,\quad N_{02}=0,\quad N_{12}=1,
$$
$$
a_{11}=0,\quad a_{12}=\frac{1}{2},
$$
$$
\overline{\omega}_G(s)=0-2+0+1.5+1.5-0.5-0.5=0.
$$
For $s=\{1,2,3,4,5\}$, we have the same $l,I_0,I_1$, 
$$
N_{01}=0,\quad N_{11}=1,\quad N_{02}=0,\quad N_{12}=1,
$$
$$
a_{11}=\frac{1}{2},\quad a_{12}=\frac{1}{2},
$$
$$
\overline{\omega}_G(s)=2-5+0+1.5+1-1-0.5=-2.
$$

\begin{figure}[H]
	\begin{center}
		\includegraphics[width=40mm]{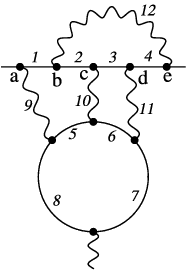}
		\caption{\label{fig_example_onloop}Example of a Feynman graph with an external photon on a lepton loop}
	\end{center}
\end{figure}

If $G$ is the graph from FIG. \ref{fig_example_onloop}, we have
$$
r=2,\quad B_1=\{\mathrm{a,c,d}\},\quad B_2=\{\mathrm{b,e}\}.
$$

If $s=\{1,2,4,5,6,7,8,9,10,11,12\}$, we have
$$
l=2,\quad I_1=\{\mathrm{a,b,c}\},\quad I_2=\{\mathrm{d,e}\},\quad I_0=\emptyset,
$$
$$
N_{01}=0,\ N_{11}=1,\ N_{21}=1,\ N_{02}=0,\ N_{12}=1,\ N_{22}=1,
$$
$$
a_{11}=a_{21}=a_{12}=a_{22}=1,
$$
$$
\overline{\omega}_G(s)=6-11+2+2.5+2-4-0.5=-3.
$$
If $s=\{1,2,4,5,6,7,8,9,11,12\}$, we have the same $l,I_0,I_1,I_2$,
$$
N_{01}=0,\ N_{11}=2,\ N_{21}=1,\ N_{02}=0,\ N_{12}=1,\ N_{22}=1,
$$
$a_{jb}$ are the same,
$$
\overline{\omega}_G(s)=4-10+2+2.5+2.5-4-0.5=-3.5.
$$
If $s=\{2,3,4,12\}$, we have
$$
l=1,\quad I_1=\{\mathrm{b,c,d,e}\},\quad I_0=\{\mathrm{a}\},
$$
$$
N_{01}=1,\ N_{11}=2,\ N_{02}=0,\ N_{12}=1,
$$
$$
a_{11}=1,\quad a_{12}=\frac{1}{2},
$$
$$
\overline{\omega}_G(s)=2-4+0+2+1.5-1.5-0.5=-0.5.
$$
If $s=\{2,3,4,6,10,11\}$, we have the same $l,I_0,I_1$,
$$
N_{01}=1,\ N_{11}=1,\ N_{02}=0,\ N_{12}=2,
$$
$$
a_{11}=1,\quad a_{12}=0,
$$
$$
\overline{\omega}_G(s)=2-6+0.5+2+1.5-1-0.5=-1.5.
$$
If $s=\{2,3,4,5,6,9,11\}$, we have the same $l,I_0,I_1$,
$$
N_{01}=1,\ N_{11}=2,\ N_{02}=0,\ N_{12}=2,
$$
the same $a_{jb}$,
$$
\overline{\omega}_G(s)=0-7+1+2+2-1-0.5=-3.5.
$$
If $s=\{1,2,3,4,5,6,7,8\}$, we have
$$
l=1,\quad,I_1=\{\mathrm{a,b,c,d,e}\},\quad I_0=\emptyset,
$$
$$
N_{01}=0,\ N_{11}=3,\ N_{02}=0,\ N_{12}=2,
$$
$$
a_{11}=a_{12}=0,
$$
$$
\overline{\omega}_G(s)=2-8+2+2.5+2.5-0-0.5=0.5.
$$
If $s=\{1,2,3,4,5,6,7,8,12\}$, we have the same $l,I_0,I_1$,
$$
N_{01}=0,\ N_{11}=3,\ N_{02}=0,\ N_{12}=1,
$$
$$
a_{11}=0,\quad a_{12}=\frac{1}{2},
$$
$$
\overline{\omega}_G(s)=4-9+2+2.5+2-0.5-0.5=0.5.
$$

\begin{figure}[H]
	\begin{center}
		\includegraphics[width=60mm]{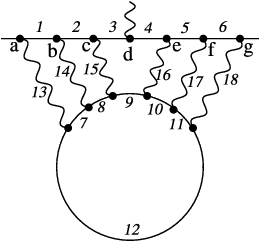}
		\caption{\label{fig_example_6ph}Example of a Feynman graph with a lepton loop of length 6}
	\end{center}
\end{figure}

If $G$ is the graph from FIG. \ref{fig_example_6ph}, we have
$$
r=2,\quad B_1=\{\mathrm{a,b,c,e,f,g}\},\quad B_2=\{\mathrm{d}\}.
$$

If $s=\{1,2,3,4,5,6,7,8,9,10,11,12\}$, we have
$$
l=1,\quad I_1=\{\mathrm{a,b,c,d,e,f,g}\},\quad I_0=\emptyset,
$$
$$
N_{01}=0,\ N_{11}=6,\ N_{02}=0,\ N_{12}=1,
$$
$$
a_{11}=0,\quad a_{12}=\frac{1}{2},
$$
$$
\overline{\omega}_G(s)=2-12+3+3.5+3.5-0.5-0.5=-1.
$$

The sequence $(l_1,G_1,l_2,\ldots,G_n,l_{n+1})$, where $n\geq 1$, $G_1,\ldots,G_n$ are self-energy subgraphs of $G$, $l_1,\ldots,l_{n+1}\in \edge[G]$, is \textbf{called} a \emph{chain} in $G$, if the following conditions are satisfied:
\begin{itemize}
\item $\vertex[G_1],\ldots,\vertex[G_n]$ do not intersect;
\item $l_1,\ldots,l_{n+1}$ are pairwise different;
\item for all $i$ ($1\leq i\leq n$), each of the lines $l_i$, $l_{i+1}$ is incident to at least one vertex of $\vertex[G_i]$;
\item the set $\{l_1,\ldots,l_{n+1}\}$ is maximal with respect to inclusion (among all the sets obtained from the objects satisfying the conditions above).
\end{itemize}
Chains can be \emph{leptonic} or \emph{photonic} (according to the type of $l_1,\ldots,l_{n+1}$). Analogously, the chain lies on $\lpath[G]$, if $l_1,\ldots,l_{n+1}\in\lpath[G]$. The chains that are obtained from each other by reversing the order are considered as the same.

\begin{figure}[H]
	\begin{center}
		\includegraphics[width=80mm]{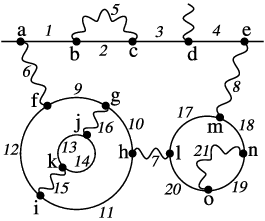}
		\caption{\label{fig_example_chains}Example of a Feynman graph with chains}
	\end{center}
\end{figure}

For example, we have the following chains in the graph $G$ from FIG. \ref{fig_example_chains}:
$$
(6,G_2,7,G_4,8),\quad (15,G_3,16),\quad (1,G_1,3),\quad (18,G_5,20),
$$
where
\begin{equation}\label{eq_example_chains_graphs}
G_1=\mathrm{bc},\ G_2=\mathrm{fghijk},\ G_3=\mathrm{jk},\ G_4=\mathrm{lmno},\ G_5=\mathrm{no}.
\end{equation}

If $G$ is a Feynman graph, $F\in\forests[G]$, $G'\in F$, by $\chains[G',F]$ we \textbf{denote} the set of all chains $(l_1,G_1,\ldots,l_{n+1})$ in $G$ such that each $G_i$ is a child of $G'$ in $F$.

For example, for the graph $G$ in FIG. \ref{fig_example_chains} and  (\ref{eq_example_chains_graphs}), for the forest
\begin{equation}\label{eq_example_chains_forest}
F=\{G,G_1,G_2,G_3,G_4,G_4\},
\end{equation}
we have
$$
\chains[G,F]=\{(1,G_1,3),\ (6,G_2,7,G_4,8)\},
$$
$$
\chains[G_2,F]=\{(15,G_3,16)\},
$$
$$
\chains[G_4,F]=\{(18,G_5,20)\}.
$$

For a fixed Feynman graph $G$, a forest $F\in\forests[G]$, and $G'\in F$, we \textbf{define} the function $\ffch_{G',F}(s)$ for all $s\in\edge[G]$ in the following way:
$$
\ffch_{G',F}(s)=\sum_{c\in\chains[G',F]}g(c,s),
$$
where
$$
g(c,s)=\begin{cases}f(c,s),\text{ if $c$ is a fermionic chain, $c$ does not lie on $\lpath[G]$, $s$ is cyclo-complete in $G$}, \\ 
\overline{f}(c,s),\text{ if $c$ is a fermionic chain, $c$ lies on $\lpath[G]$, $s$ is cyclo-complete in $G$}, \\
\tilde{f}(c,s),\text{ if $c$ is a fermionic chain, $s$ is not cyclo-complete in $G$}, \\
p(c,s),\text{ if $c$ is a photonic chain, $s$ is cyclo-complete in $G$}, \\
\tilde{p}(c,s),\text{ if $c$ is a photonic chain, $s$ is not cyclo-complete in $G$}, \end{cases}
$$
$$
f((l_1,G_1,\ldots,l_{n+1}),s)=\begin{cases}
-\frac{n}{2},\text{ if } l_1,\ldots,l_{n+1}\in s, \\
\frac{1}{2}(h(G_1,s)+\ldots+h(G_n,s)) \text{ otherwise},
\end{cases}
$$
$$
h(G',s) = \begin{cases}
1,\text{ if } \edge[G']\subseteq s, \\
0 \text{ otherwise},
\end{cases}
$$
$$
\overline{f}((l_1,G_1,\ldots,l_{n+1}),s)=\begin{cases}
-\frac{n}{2},\text{ if } l_1,\ldots,l_{n+1}\in s, \\
0 \text{ otherwise},
\end{cases}
$$
$$
\tilde{f}((l_1,G_1,\ldots,l_{n+1}),s)=\begin{cases}
-\frac{n}{2},\text{ if } l_1,\ldots,l_{n+1}\in s, \\
\frac{1}{2}|\{l_1,\ldots,l_{n+1}\}\cap s| + h(G_1,s)+\ldots+h(G_n,s) \text{ otherwise},
\end{cases}
$$
$$
p((l_1,G_1,\ldots,l_{n+1}),s)=\begin{cases}
-n,\text{ if } l_1,\ldots,l_{n+1}\in s, \\
0 \text{ otherwise},
\end{cases}
$$
$$
\tilde{p}((l_1,G_1,\ldots,l_{n+1}),s)=\begin{cases}
-n,\text{ if } l_1,\ldots,l_{n+1}\in s, \\
h(G_1,s)+\ldots+h(G_n,s) \text{ otherwise}.
\end{cases}
$$
The auxiliary functions $f,\overline{f},\tilde{f},p,\tilde{p}$ are illustrated by the following examples concerning the graph $G$ from FIG. \ref{fig_example_chains} and the subgraphs \ref{eq_example_chains_graphs}:
$$
f((1,G_1,3),\{1,2,3\})=-0.5,\quad f((1,G_1,3),\{1,2\})=0,
$$
$$
f((1,G_1,3),\{1,2,5\})=0.5,
$$
$$
\overline{f}((1,G_1,3),\{1,2,3\})=-0.5,\quad \overline{f}((1,G_1,3),\{1,2\})=0,
$$
$$
\overline{f}((1,G_1,3),\{1,2,5\})=0,
$$
$$
\tilde{f}((1,G_1,3),\{1,2,3\})=-0.5,\quad \tilde{f}((1,G_1,3),\{1,2\})=0.5,
$$
$$
\tilde{f}((1,G_1,3),\{1,2,5\})=1.5,
$$
$$
p((6,G_2,7,G_4,8),\{6,7,9,10,11,12,15,16\})=0,
$$
$$
\tilde{p}((6,G_2,7,G_4,8),\{6,7,9,10,11,12,15,16\})=0,
$$
$$
p((6,G_2,7,G_4,8),\{6,7,8,9,10,11,12,15,16\})=-2,
$$
$$
\tilde{p}((6,G_2,7,G_4,8),\{6,7,8,9,10,11,12,15,16\})=-2,
$$
$$
p((6,G_2,7,G_4,8),\{6,7,9,10,11,12,13,14,15,16\})=0,
$$
$$
\tilde{p}((6,G_2,7,G_4,8),\{6,7,9,10,11,12,13,14,15,16\})=1.
$$
The set $s=\{1,3,9,10,\ldots,20\}$ is cyclo-complete in $G$. If (\ref{eq_example_chains_forest}) is satisfied, we have
$$
\ffch_{G,F}(s) = \overline{f}((1,G_1,3),s) + p((6,G_2,7,G_4,8),s) 
$$
$$
= -0.5 + 0 = -0.5,
$$
$$
\ffch_{G_2,F}(s) = p((15,G_3,16),s) = -1,
$$
$$
\ffch_{G_4,F}(s) = f((18,G_5,20),s) = -0.5.
$$
For the not cyclo-complete set $s=\{13,14,19,20\}$, we have
$$
\ffch_{G,F}(s) = \tilde{f}((1,G_1,3),s) + \tilde{p}((6,G_2,7,G_4,8),s) 
$$
$$
= 0 + 0 = 0,
$$
$$
\ffch_{G_2,F}(s) = \tilde{p}((15,G_3,16),s) = 1,
$$
$$
\ffch_{G_4,F}(s) = \tilde{f}((18,G_5,20),s) = 0.5.
$$
For another not cyclo-complete set $s=\{2,5,6,8,9,10,11,12,17,18,19,20\}$,
we have
$$
\ffch_{G,F}(s) = \tilde{f}((1,G_1,3),s) + \tilde{p}((6,G_2,7,G_4,8),s) 
$$
$$
= 1 + 0 = 1.
$$

Suppose the Feynman graph $G$ is fixed. Let us \textbf{define} the functions $\tilde{\omega}_{G',F}(s)$, $\tilde{\omega}'_{G',F}(s)$, where $F\in\forests[G]$, $G'\in F$, $s\subseteq \edge[G]$:
$$
\tilde{\omega}_{G',F}(s)=\begin{cases}
\max(\overline{\omega}_{G'/F}(s),\omega_{G'/F}(s)) - \ffch_{G',F}(s),\text{ if $s$ is cyclo-complete in $G$, and $G'$ lies on $\lpath[G]$}, \\
\omega_{G'/F}(s) - \ffch_{G',F}(s) \text{ otherwise},
\end{cases}
$$
$$
\tilde{\omega}'_{G',F}(s) = \min(0,\tilde{\omega}_{G',F}(s)+\csub[G',s]),
$$
where
$$
\csub[G',s] = \begin{cases}
\csubfseop(s),\text{ if $G'$ is a lepton self-energy subgraph, $G'$ lies on $\lpath[G]$}, \\
\csubfseol(s),\text{ if $G'$ is a lepton self-energy subgraph, $G'$ does not lie on $\lpath[G]$}, \\
\csubi(s),\text{ if $G'$ is vertexlike, $G'$ lies on $\lpath[G]$,} \\
\quad\quad\quad\text{the external photon of $G$ is incident to some $v\in \vertex[G']$}, \\
\csubvop(s),\text{ if $G'$ is vertexlike, $G'$ lies on $\lpath[G]$}, \\ \quad\quad\quad \text{there is no $v\in\vertex[G']$ incident to the external photon of $G$ }, \\
\csubvol(s) \text{ if $G'$ is vertexlike, $G'$ does not lie on $\lpath[G]$}, \\
0,\text{ in the other cases},
\end{cases}
$$
the values $\csubfseop(s)$, $\csubfseol(s)$, $\csubi(s)$, $\csubvop(s)$, $\csubvol(s)$ are taken from Table \ref{table_carlo_constants} depending on whether $s$ is cyclo-complete in $G$ and whether the external photon of $G$ is incident to $v\in\lpath[G]$. The table constants were obtained by numerical experiments with the Monte Carlo convergence speed at the 4-loop level.

\begin{table}[H]
\caption{\label{table_carlo_constants}The constants that are used for obtaining $\ffdeg_0(s)$. The values depend on whether $s$ is cyclo-complete in the whole graph $G$ and whether the external photon of $G$ is incident to $v\in\lpath[G]$ (``on path'') or not (``on loop'').}
\begin{ruledtabular}
\begin{tabular}{lllll}
Constant & \multicolumn{2}{c}{cyclo-complete} & \multicolumn{2}{c}{not cyclo-complete} \\
& on path & on loop & on path & on loop \\
\colrule 
$\csubfseop(s)$ & 0 & 0 & 0.293 & 0 \\
$\csubfseol(s)$ & 0.021 & 0 & 0 & 0 \\
$\csubi(s)$ & 0 & 0.011 & 0.228 & 0.622 \\
$\csubvop(s)$ & 0.085 & 0 & 0.041 & 0.394 \\
$\csubvol(s)$ & 0.185 & 0 & 0 & 0 \\
$\cbigz$ & 0.695 & 1.079 & ... & ... \\
$\cbigf$ & 0.285 & 0.38 & ... & ... \\
$\cadd(s)$ & 0.199 & -0.122 & 0 & 0.187 \\
$\csat(s)$ & 0.46 & 0.716 & 0.785 & 0.684
\end{tabular}
\end{ruledtabular}
\end{table}

Let us examine several examples. Suppose the graph $G$ is taken from FIG. \ref{fig_example_chains}, the conditions (\ref{eq_example_chains_graphs}) and (\ref{eq_example_chains_forest}) are satisfied. If $s=\{13,14,19,20\}$, we have for 
$$
\tilde{\omega}_{G,F}(s)=\omega_{G,F}(s)-\ffch_{G,F}(s)=0-0=0,
$$
$$
\tilde{\omega}'_{G,F}(s)=\min(0,0+\csubi(s))=\min(0,\csubi(s)),
$$
$$
\tilde{\omega}_{G_1,F}(s)=\omega_{G_1,F}(s)-\ffch_{G_1,F}(s)=0-0=0,
$$
$$
\tilde{\omega}'_{G_1,F}(s)=\min(0,\csubi(s))=\min(0,\csubfseop(s)),
$$
$$
\tilde{\omega}_{G_2,F}(s)=\omega_{G_2,F}(s)-\ffch_{G_2,F}(s)=0-1=-1,
$$
$$
\tilde{\omega}'_{G_2,F}(s)=\min(0,-1)=-1,
$$
$$
\tilde{\omega}_{G_4,F}(s)=\omega_{G_4,F}(s)-\ffch_{G_4,F}(s)=-0.5-0.5=-1,
$$
$$
\tilde{\omega}'_{G_4,F}(s)=\min(0,-1)=-1,
$$
$$
\tilde{\omega}_{G_5,F}(s)=\omega_{G_5,F}(s)-\ffch_{G_5,F}(s)=-0.5-0=-0.5,
$$
$$
\tilde{\omega}'_{G_5,F}(s)=\min(0,-0.5+\csubfseol(s)).
$$
If $F=\{G,G_1,G_2,G_3,G_4,G_5,G_6\}$ with $G_1,\ldots,G_5$ from (\ref{eq_example_chains_graphs}) and $G_6=\mathrm{ghijk}$, $s=\{9,12,15,16\}$, we have
$$
\tilde{\omega}_{G_6,F}(s) = \omega_{G_6,F}(s) - \ffch_{G_6,F}(s) = -2-(-1) = -1,
$$
$$
\tilde{\omega}'_{G_6,F}(s)=\min(0,-1+\csubvol(s)).
$$
Suppose $G$ is from FIG. \ref{fig_example_noloops},
\begin{equation}\label{eq_example_noloops_forests}
F_1=\{G,G_1,G_2,G_3\},\quad F_2=\{G,G_1,G_2,G_4\},
\end{equation}
\begin{equation}\label{eq_example_noloops_graphs}
G_1=\mathrm{bcdefgh},\ G_2=\mathrm{cdef},\ G_3=\mathrm{cde},\ G_4=\mathrm{def}.
\end{equation}
If $s=\{4,5,10,11\}$, we have
$$
\tilde{\omega}_{G,F_1}(s)=\tilde{\omega}_{G,F_2}(s)=0,
$$
$$
\tilde{\omega}'_{G,F_1}(s)=\tilde{\omega}'_{G,F_2}(s) = \min(0,\csubi(s)),
$$
$$
\tilde{\omega}_{G_1,F_1}(s)=\tilde{\omega}_{G_1,F_2}(s)=\max(\overline{\omega}_{G_1,F_1}(s),\omega_{G_1,F_1}(s))
$$
$$
= \max(-1.5,-1)=-1,
$$
$$
\tilde{\omega}'_{G_1,F_1}(s)=\tilde{\omega}'_{G_1,F_2}(s) = \min(0,-1+\csubi(s)),
$$
$$
\tilde{\omega}_{G_2,F_1}(s)=\max(\overline{\omega}_{G_2,F_1}(s),\omega_{G_2,F_1}(s)) = \max(0.5,-0.5)=0.5,
$$
$$
\tilde{\omega}'_{G_2,F_1}(s)=\min(0,0.5+\csubfseop(s)),
$$
$$
\tilde{\omega}_{G_2,F_2}(s)=\max(\overline{\omega}_{G_2,F_2}(s),\omega_{G_2,F_2}(s)) = \max(-0.5,0)=0,
$$
$$
\tilde{\omega}'_{G_2,F_2}(s)=\min(0,\csubfseop(s)),
$$
$$
\tilde{\omega}_{G_3,F_1}(s)=\max(\overline{\omega}_{G_3,F_1}(s),\omega_{G_3,F_1}(s)) = \max(-1,-0.5)=-0.5,
$$
$$
\tilde{\omega}'_{G_3,F_1}(s)=\min(0,-0.5+\csubvop(s)),
$$
$$
\tilde{\omega}_{G_4,F_2}(s)=\max(\overline{\omega}_{G_4,F_2}(s),\omega_{G_4,F_2}(s)) = \max(0,0)=0,
$$
$$
\tilde{\omega}'_{G_4,F_2}(s)=\min(0,\csubvop(s)).
$$
Suppose $G$ is taken from FIG. \ref{fig_example_6ph}, 
$$
F=\{G\},\quad s=\{1,2,\ldots,12\}.
$$
We have 
$$
\tilde{\omega}_{G,F}(s) = \max(\overline{\omega}_{G,F}(s),\omega_{G,F}(s)) = \max(-1,-4)=-1,
$$
$$
\tilde{\omega}'_{G,F}(s) = \min(0,-1+\csubi(s)).
$$

\subsubsection{The formula}

Suppose $G$ is a vertexlike Feynman graph.

By $\forestsmax[G]$ we denote the set of all maximal sets from $\forests[G]$ (with respect to inclusion).

By definition, put
$$
\ffdeg_0(s) = \begin{cases}
\cbigz+(\cbigf-\cbigz)\frac{|\photon(s)|}{|\photon(\edge[G])|},\text{ if $\lepton(\edge[G])\subseteq s$, and} \\
\quad\quad\quad \text{there exists $F\in\forestsmax[G]$ such that $\tilde{\omega}_{G',F}(s)\geq 0$ for all $G'\in F$}, \\
\min_{F\in\forestsmax[G]} D_F(s) \text{ otherwise},
\end{cases}
$$
where 
$$
D_F(s)=\tau\left(-\sum_{G'\in F} \tilde{\omega}'_{G',F}(s),\ \cadd(s),\ \csat(s) \right),
$$
$\cbigz$, $\cbigf$, $\cadd(s)$, $\csat(s)$ are taken from Table \ref{table_carlo_constants},
$$
\tau(x,C_{\mathrm{a}},C_{\mathrm{s}}) = \frac{1}{2} (C_{\mathrm{a}}+C_{\mathrm{s}}) \times \mu\left( \frac{x-C_{\mathrm{s}}}{C_{\mathrm{a}}+C_{\mathrm{s}}}\right),
$$
$$
\mu(t)=2+t+\sqrt{t^2+0.25}.
$$

Here, the table constants were determined by numerical experiments with the Monte Carlo convergence speed at the 4-loop level; $\mu(t)$ is a smooth approximation for $2+\max(0,2t)$; the function $\tau(x,C_{\mathrm{a}},C_{\mathrm{s}})$ is a smooth approximation for $\max(x,C_{\mathrm{s}})+C_{\mathrm{a}}$.

Let us examine several examples. Suppose $G$ is Feynman graph from FIG. \ref{fig_example_noloops}. In this case,
$$
\forestsmax[G]=\{F_1,F_2\},
$$
where (\ref{eq_example_noloops_forests}) and (\ref{eq_example_noloops_graphs}) are satisfied. If $s=\{4,5,10,11\}$, we have 
$$
\ffdeg_0(s)=\min\left(\tau(-\tilde{\omega}'_{G,F_1}(s)-\tilde{\omega}'_{G_1,F_1}(s) -\tilde{\omega}'_{G_2,F_1}(s) -\tilde{\omega}'_{G_3,F_1}(s),\cadd(s),\csat(s)), \right.
$$
$$
\left. \tau(-\tilde{\omega}'_{G,F_2}(s)-\tilde{\omega}'_{G_1,F_2}(s) -\tilde{\omega}'_{G_2,F_2}(s) -\tilde{\omega}'_{G_4,F_2}(s),\cadd(s),\csat(s))\right)
$$
Since all the constants $\csubfseop$, $\csubfseol$, $\csubi$, $\csubvop$, $\csubvol$ are nonnegative, we have
$$
\ffdeg_0(s)=\tau(\max(0,1-\csubi(s)), \cadd(s),\csat(s)).
$$
If $s=\{4,5,10\}$, we have
$$
\tilde{\omega}_{G,F_2}(s)= \tilde{\omega}_{G_1,F_2}(s) = \tilde{\omega}_{G_2,F_2}(s) = \tilde{\omega}_{G_4,F_2}(s)=0,
$$
but we still use $D_F$, because $s$ does not contain $\lepton(\edge[G])$:
$$
\ffdeg_0(s)=D_{F_2}(s)=\tau(0, \cadd(s), \csat(s)).
$$
If $s=\{1,2,3,4,5,6,7,8,10\}$, we have
$$
\tilde{w}_{G,F_2}(s)=0,\ \tilde{w}_{G_1,F_2}(s)=0,\ \tilde{w}_{G_2,F_2}(s)=0.5,\ \tilde{w}_{G_4,F_2}(s)=0, 
$$
$$
\ffdeg_0(s) = \frac{1}{4} \cbigf + \frac{3}{4} \cbigz.
$$

Suppose $G$ is Feynman graph from FIG. \ref{fig_example_6ph}. In this case, we have
$$
\forestsmax[G]=\{F\},\quad F=\{G\}.
$$
If $s=\{1,2,\ldots,12\}$, we have $\lepton(\edge[G])\subseteq s$, but we use $D_F$, because $\tilde{\omega}_{G,F}(s)=-1<0$:
$$
\ffdeg_0(s)=\tau(\max(0, \csubi(s)-1), \cadd(s), \csat(s)).
$$

\subsection{Additional stabilization techniques}

During the integration, we perform the summation of the values $I(\underline{z})/g(\underline{z})$, where $I$ is the integrand, $g$ is the probability density function. Since these values are unbounded, it is very important to prevent from occasional emergence of very large values, but to allow their systematic occurence; we used the technique described in Section III.D of \cite{volkov_prd} adapted to GPUs. The presence of acute peaks in the integrands can lead to an error underestimation; to prevent from this, we used the heuristic described in Section IV.F of \cite{volkov_gpu}.

\section{TECHNICAL INFORMATION AND NUMERICAL CANCELLATIONS}\label{sec_technical}

In our approach, all divergences are numerically cancelled at the level of the integrands. This cancellation leads to round-off errors. We evaluate the integrand values using  interval arithmetic to control the errors. The samples that result in too large intervals are evaluated with increased precisions. The algorithm of  interval acceptance and rejection was described in detail in \cite{volkov_5loops_2019} (Section IV.A). It is constructed so that the round-off error is small relative to the Monte Carlo statistical error even in the worst case and takes into account that the round-off error can have a non-zero mean value and we cannot add them as statistical errors. In this calculation, it was modified so that it also provides correct values for the needed graph subsets.

The integrand expressions at the 5-loop level are huge. Special tricks are required to evaluate them on GPUs. The general scheme of the realization as well as the interval arithmetic are described in \cite{volkov_5loops_2019}. We have generated the integrand code for different precisions separately:
\begin{itemize}
\item ``Eliminated Interval Arithmetic'' (EIA) is an approach that works with the machine \texttt{double} precision, but significantly faster than the usual interval arithmetic (the price for this is larger intervals). It is described in Section IV.C of \cite{volkov_gpu}.
\item The usual machine \texttt{double} precision interval arithmetic.
\item Arbitrary-precision interval arithmetic with the mantissa size as a parameter.
\end{itemize}

The codegenerator for the integrands was written in the D programming language. The generation took a couple of months on ITP KIT office computers. The code was generated in C++\footnote{Formally it is C++, but for the most part the constructions from C were used.} with CUDA, the size is 136 GB and 294 GB for graphs with lepton loops and without lepton loops, respectively. The size in the compiled form is 191 GB and 393 GB, respectively. The compilation with \texttt{nvcc} was performed on the supercomputer ``HoreKa'' (Karlsruhe, Germany) with the help of the processors Intel Xeon Platinum 8368 (38 cores) and took about 1 core-month.

We evaluate all Feynman graphs directly; we never use known lower-order formulas for the vacuum polarization. These substitutions could significantly improve the precision for classes such as Set I(a), but the impact on the overall accuracy is negligible, and the individual precision achieved is enough for verification. For example, it is dominated by Set V that has no vacuum polarizations at all\footnote{There are many graphs that have only one 1-loop vacuum polarization insertion. They provide a significant contribution (if we ignore Set V), but we do not believe that using explicit one-loop formulas would significantly improve the speed and precision: The dimensionality of the integrand would be reduced by 2 (from 13 to 11), but the integrand structure would become more complicated.}. On the other hand, the use of explicit formulas makes the code more complicated and unreliable.

Table \ref{table_round_off_contr} contains the information about the contributions of the Monte Carlo samples that required increased precisions. $\triangle^{\text{fail}}_{\text{EIA}}$, $\triangle^{\text{fail}}_{\text{IA}}$, $\triangle^{\text{fail}}_{\text{128}}$, $\triangle^{\text{fail}}_{\text{192}}$ are the contributions of the samples for which the following precision failed: Eliminated Interval Arithmetic, the usual mashine \texttt{double} precision interval arithmetic, 128-bit-mantissa interval arithmetic, 192-bit-mantissa interval arithmetic, respectively. The integrands and the probability density functions tend very quickly towards infinity at the boundary; sometimes the situation arises that the probability density value is too large to be stored in the mashine \texttt{double} precision. This is not a problem for our integration program, but the contribution $\triangle^{\text{dens}}_{> \text{double}}$ of these samples is also given in the table.

The corresponding numbers of the Monte Carlo samples are given in Table \ref{table_round_off_np}. The table also includes the number $N^{\text{fail}}_{\text{256}}$ of the samples for which the 256-bit-mantissa precision failed. It is the highest precision in our integration, these samples are considered as giving $0$.

Table \ref{table_oscillations} contains an information about oscillations in individual graph contributions and in integrands. This concerns only the new calculation. The part referring to the graphs without lepton loops can be compared with the corresponding table in \cite{volkov_5loops_2019}. The use of $M^2=-m^2$ instead of $M^2=m^2$ in (\ref{eq_u_vertex}) and (\ref{eq_u_self_energy}) reduced the oscillations by about $1.5$ times.

\begin{longtable}{ccccccc}\caption{Contributions of the Monte-Carlo samples with round-off errors and with the probability density out of mashine \texttt{double} (continued)} \\
\hline \hline Class & Value & $\triangle^{\text{fail}}_{\text{EIA}}$ & $\triangle^{\text{fail}}_{\text{IA}}$ & $\triangle^{\text{fail}}_{\text{128}}$ & $\triangle^{\text{fail}}_{\text{192}}$ & $\triangle^{\text{dens}}_{> \text{double}}$ \\ \hline  \endhead
\caption{Contributions of the Monte-Carlo samples with round-off errors and with the probability density out of mashine \texttt{double}}\label{table_round_off_contr} \\
\hline \hline Class & Value & $\triangle^{\text{fail}}_{\text{EIA}}$ & $\triangle^{\text{fail}}_{\text{IA}}$ & $\triangle^{\text{fail}}_{\text{128}}$ & $\triangle^{\text{fail}}_{\text{192}}$ & $\triangle^{\text{dens}}_{> \text{double}}$ \\ \hline  \endfirsthead
\hline \endfoot  \hline \hline \endlastfoot
I(a) & 0.00047105(16) & $-0.00069$ & $2.3\times 10^{-5}$ & $5.7\times 10^{-14}$ & $3.0\times 10^{-25}$ & $-2.1\times 10^{-34}$ \\
I(b) & 0.0070081(12) & $-0.0013$ & $0.00058$ & $-9.2\times 10^{-13}$ & $1.0\times 10^{-23}$ & $1.1\times 10^{-34}$ \\
I(c) & 0.0234643(26) & $0.0074$ & $0.0029$ & $7.3\times 10^{-12}$ & $-3.0\times 10^{-24}$ & $-1.4\times 10^{-33}$ \\
I(d) & 0.00380370(61) & $0.00076$ & $0.00071$ & $1.2\times 10^{-9}$ & $-6.6\times 10^{-15}$ & $-2.9\times 10^{-26}$ \\
I(e) & 0.010289(11) & $0.0051$ & $0.0018$ & $6.8\times 10^{-7}$ & $1.1\times 10^{-17}$ & $-7.8\times 10^{-20}$ \\
I(f) & 0.00757106(47) & $0.0045$ & $0.0019$ & $5.8\times 10^{-8}$ & $3.3\times 10^{-14}$ & $-4.0\times 10^{-25}$ \\
I(g) & 0.0285696(18) & $0.014$ & $0.0051$ & $2.4\times 10^{-8}$ & $-1.9\times 10^{-14}$ & $-5.9\times 10^{-20}$ \\
I(h) & 0.0016826(63) & $0.0016$ & $-0.00067$ & $-2.9\times 10^{-7}$ & $2.9\times 10^{-13}$ & $1.8\times 10^{-23}$ \\
I(i) & 0.01726(29) & $-0.010$ & $-0.0050$ & $-9.9\times 10^{-5}$ & $-1.9\times 10^{-8}$ & $9.6\times 10^{-15}$ \\
I(j) & 0.0004038(63) & $0.0037$ & $-0.0014$ & $-1.1\times 10^{-9}$ & $-1.2\times 10^{-17}$ & $2.2\times 10^{-22}$ \\
II(a) & -0.1094945(50) & $-0.0070$ & $-0.0012$ & $1.4\times 10^{-9}$ & $-1.4\times 10^{-21}$ & $5.0\times 10^{-16}$ \\
II(b) & -0.473625(27) & $-0.052$ & $-0.0082$ & $4.5\times 10^{-8}$ & $-5.0\times 10^{-14}$ & $-1.6\times 10^{-12}$ \\
II(c) & -0.116506(12) & $-0.041$ & $-0.0052$ & $6.3\times 10^{-8}$ & $-1.1\times 10^{-19}$ & $3.2\times 10^{-13}$ \\
II(d) & -0.24291(15) & $-0.042$ & $-0.0078$ & $4.9\times 10^{-6}$ & $-3.1\times 10^{-16}$ & $-3.6\times 10^{-14}$ \\
II(e) & -1.34235(54) & $-0.77$ & $-0.014$ & $0.00014$ & $3.9\times 10^{-8}$ & $-6.0\times 10^{-13}$ \\
II(f) & -2.43553(30) & $-1.8$ & $0.097$ & $0.00095$ & $3.4\times 10^{-7}$ & $-2.5\times 10^{-14}$ \\
III(a) & 2.12726(14) & $0.68$ & $-0.11$ & $-0.00038$ & $-1.6\times 10^{-6}$ & $-1.1\times 10^{-6}$ \\
III(b) & 3.32730(32) & $0.99$ & $-0.20$ & $-0.00026$ & $-1.1\times 10^{-6}$ & $-5.5\times 10^{-7}$ \\
III(c) & 4.9199(15) & $4.9$ & $1.3$ & $0.0010$ & $4.6\times 10^{-7}$ & $-1.2\times 10^{-11}$ \\
IV & -7.7303(16) & $-4.1$ & $0.81$ & $0.0056$ & $-3.6\times 10^{-5}$ & $-5.7\times 10^{-6}$ \\
V & 6.857(81) & $11.4$ & $2.7$ & $0.12$ & $-0.00053$ & $1.9\times 10^{-10}$ \\
VI(a) & 1.041537(82) & $0.45$ & $-0.078$ & $-0.00060$ & $5.2\times 10^{-8}$ & $1.9\times 10^{-16}$ \\
VI(b) & 1.34697(11) & $0.26$ & $-0.17$ & $-0.0013$ & $-1.1\times 10^{-6}$ & $1.3\times 10^{-13}$ \\
VI(c) & -2.53312(49) & $-1.2$ & $0.039$ & $0.0032$ & $3.9\times 10^{-6}$ & $-7.2\times 10^{-13}$ \\
VI(d) & 1.8468(22) & $1.5$ & $1.6$ & $-0.013$ & $-1.5\times 10^{-5}$ & $-1.0\times 10^{-11}$ \\
VI(e) & -0.43129(17) & $-0.28$ & $-0.00087$ & $1.1\times 10^{-7}$ & $-3.1\times 10^{-16}$ & $3.2\times 10^{-17}$ \\
VI(f) & 0.77154(23) & $0.59$ & $-0.11$ & $-0.0011$ & $-5.1\times 10^{-7}$ & $-1.6\times 10^{-13}$ \\
VI(g) & -1.5965(10) & $-1.4$ & $-0.018$ & $-0.0056$ & $-1.0\times 10^{-5}$ & $-3.6\times 10^{-12}$ \\
VI(h) & 0.18554(68) & $0.070$ & $-0.0019$ & $0.00017$ & $2.1\times 10^{-7}$ & $-6.6\times 10^{-13}$ \\
VI(i) & -0.04396(10) & $-0.13$ & $-0.038$ & $-3.6\times 10^{-5}$ & $2.2\times 10^{-9}$ & $-5.1\times 10^{-14}$ \\
VI(j) & -0.22920(43) & $0.032$ & $-0.083$ & $-0.00064$ & $-6.0\times 10^{-7}$ & $1.7\times 10^{-12}$ \\
VI(k) & 0.67974(39) & $0.43$ & $0.80$ & $0.00064$ & $7.9\times 10^{-9}$ & $-2.2\times 10^{-13}$ \\
\end{longtable}

\begin{longtable}{cccccccc}\caption{Statistics of round-off errors in the integrand evaluations and the Monte Carlo samples with the probability density out of mashine \texttt{double} (continued)} \\
\hline \hline Class & $N_{\text{total}}$ & $N^{\text{fail}}_{\text{EIA}}$ & $N^{\text{fail}}_{\text{IA}}$ & $N^{\text{fail}}_{\text{128}}$ & $N^{\text{fail}}_{\text{192}}$ & $N^{\text{fail}}_{\text{256}}$ & $N^{\text{dens}}_{> \text{double}}$ \\ \hline  \endhead
\caption{Statistics of round-off errors in the integrand evaluations and the Monte Carlo samples with the probability density out of mashine \texttt{double}}\label{table_round_off_np} \\
\hline \hline Class & $N_{\text{total}}$ & $N^{\text{fail}}_{\text{EIA}}$ & $N^{\text{fail}}_{\text{IA}}$ & $N^{\text{fail}}_{\text{128}}$ & $N^{\text{fail}}_{\text{192}}$ & $N^{\text{fail}}_{\text{256}}$ & $N^{\text{dens}}_{> \text{double}}$ \\ \hline  \endfirsthead
\hline \endfoot  \hline \hline \endlastfoot
I(a) & $15\times 10^{11}$ & $38\times 10^{10}$ & $15\times 10^{9}$ & $19\times 10^{7}$ & $36\times 10^{4}$ & $3620$ & $5$ \\
I(b) & $24\times 10^{11}$ & $54\times 10^{10}$ & $40\times 10^{9}$ & $25\times 10^{7}$ & $15\times 10^{4}$ & $1171$ & $6$ \\
I(c) & $81\times 10^{10}$ & $18\times 10^{10}$ & $23\times 10^{9}$ & $22\times 10^{6}$ & $3037$ & $8$ & $6$ \\
I(d) & $10^{12}$ & $19\times 10^{10}$ & $31\times 10^{9}$ & $21\times 10^{6}$ & $5452$ & $11$ & $85$ \\
I(e) & $29\times 10^{11}$ & $51\times 10^{10}$ & $69\times 10^{9}$ & $77\times 10^{6}$ & $42129$ & $114$ & $57$ \\
I(f) & $84\times 10^{10}$ & $11\times 10^{10}$ & $19\times 10^{9}$ & $96\times 10^{5}$ & $4829$ & $8$ & $99$ \\
I(g) & $15\times 10^{11}$ & $25\times 10^{10}$ & $46\times 10^{9}$ & $16\times 10^{6}$ & $5962$ & $2$ & $1057$ \\
I(h) & $50\times 10^{11}$ & $61\times 10^{10}$ & $58\times 10^{9}$ & $87\times 10^{5}$ & $3513$ & $3$ & $3040$ \\
I(i) & $85\times 10^{12}$ & $10^{13}$ & $12\times 10^{11}$ & $23\times 10^{7}$ & $13\times 10^{4}$ & $199$ & $19304$ \\
I(j) & $74\times 10^{10}$ & $25\times 10^{10}$ & $46\times 10^{9}$ & $43\times 10^{7}$ & $19\times 10^{5}$ & $25355$ & $3096$ \\
II(a) & $49\times 10^{11}$ & $12\times 10^{11}$ & $95\times 10^{9}$ & $72\times 10^{7}$ & $23\times 10^{5}$ & $55926$ & $6528$ \\
II(b) & $65\times 10^{11}$ & $18\times 10^{11}$ & $13\times 10^{10}$ & $29\times 10^{7}$ & $50\times 10^{4}$ & $3981$ & $22909$ \\
II(c) & $26\times 10^{11}$ & $61\times 10^{10}$ & $54\times 10^{9}$ & $87\times 10^{5}$ & $682$ & $0$ & $23303$ \\
II(d) & $14\times 10^{12}$ & $29\times 10^{11}$ & $11\times 10^{10}$ & $24\times 10^{6}$ & $4433$ & $3$ & $52772$ \\
II(e) & $27\times 10^{11}$ & $13\times 10^{11}$ & $69\times 10^{9}$ & $68\times 10^{7}$ & $27\times 10^{5}$ & $20662$ & $11\times 10^{4}$ \\
II(f) & $85\times 10^{10}$ & $52\times 10^{10}$ & $43\times 10^{9}$ & $28\times 10^{7}$ & $11\times 10^{5}$ & $10002$ & $27199$ \\
III(a) & $10^{13}$ & $44\times 10^{11}$ & $24\times 10^{10}$ & $12\times 10^{8}$ & $42\times 10^{5}$ & $45636$ & $22\times 10^{4}$ \\
III(b) & $15\times 10^{12}$ & $54\times 10^{11}$ & $38\times 10^{10}$ & $17\times 10^{8}$ & $57\times 10^{5}$ & $41232$ & $54\times 10^{4}$ \\
III(c) & $11\times 10^{12}$ & $63\times 10^{11}$ & $56\times 10^{10}$ & $77\times 10^{8}$ & $43\times 10^{6}$ & $49\times 10^{4}$ & $13\times 10^{5}$ \\
IV & $78\times 10^{12}$ & $35\times 10^{12}$ & $29\times 10^{11}$ & $21\times 10^{9}$ & $10^{8}$ & $13\times 10^{5}$ & $63\times 10^{5}$ \\
V & $24\times 10^{13}$ & $51\times 10^{12}$ & $90\times 10^{11}$ & $19\times 10^{10}$ & $10^{9}$ & $50\times 10^{5}$ & $45\times 10^{5}$ \\
VI(a) & $41\times 10^{10}$ & $21\times 10^{10}$ & $25\times 10^{9}$ & $47\times 10^{6}$ & $46527$ & $457$ & $4932$ \\
VI(b) & $10^{12}$ & $44\times 10^{10}$ & $36\times 10^{9}$ & $70\times 10^{6}$ & $13\times 10^{4}$ & $1745$ & $30089$ \\
VI(c) & $30\times 10^{11}$ & $21\times 10^{11}$ & $11\times 10^{10}$ & $20\times 10^{7}$ & $11\times 10^{5}$ & $23583$ & $40\times 10^{4}$ \\
VI(d) & $41\times 10^{12}$ & $26\times 10^{12}$ & $12\times 10^{11}$ & $36\times 10^{8}$ & $28\times 10^{6}$ & $72\times 10^{4}$ & $10^{7}$ \\
VI(e) & $55\times 10^{10}$ & $35\times 10^{10}$ & $20\times 10^{9}$ & $28\times 10^{6}$ & $12\times 10^{4}$ & $2736$ & $37778$ \\
VI(f) & $12\times 10^{11}$ & $69\times 10^{10}$ & $48\times 10^{9}$ & $10^{8}$ & $68\times 10^{4}$ & $16916$ & $31312$ \\
VI(g) & $10^{13}$ & $60\times 10^{11}$ & $26\times 10^{10}$ & $54\times 10^{7}$ & $30\times 10^{5}$ & $65104$ & $11\times 10^{5}$ \\
VI(h) & $12\times 10^{12}$ & $40\times 10^{11}$ & $27\times 10^{10}$ & $57\times 10^{7}$ & $22\times 10^{5}$ & $50017$ & $22\times 10^{4}$ \\
VI(i) & $62\times 10^{10}$ & $18\times 10^{10}$ & $10^{10}$ & $13\times 10^{6}$ & $20159$ & $281$ & $5509$ \\
VI(j) & $63\times 10^{10}$ & $40\times 10^{10}$ & $28\times 10^{9}$ & $99\times 10^{6}$ & $61\times 10^{4}$ & $11317$ & $11062$ \\
VI(k) & $32\times 10^{11}$ & $13\times 10^{11}$ & $20\times 10^{9}$ & $26\times 10^{5}$ & $507$ & $0$ & $361$ \\
\end{longtable}

\begin{longtable}{cccccc}\caption{Oscillations in the individual graph contributions and integrals; here, $a_i=\int I_i(\underline{z})d\underline{z}$ is the contribution of the $i$-th graph to the value, $I_i$ is the corresponding Feynman parametric integrand. (continued)} \\
\hline \hline Class & Value $=\sum_i a_i$ & $\sum_i |a_i|$ & $\max_i |a_i|$ & $\sum_i \int |I_i(\underline{z})| d\underline{z}$ & $N_{\text{graph}}$ \\ \hline  \endhead
\caption{Oscillations in the individual graph contributions and integrals; here, $a_i=\int I_i(\underline{z})d\underline{z}$ is the contribution of the $i$-th graph to the value, $I_i$ is the corresponding Feynman parametric integrand.}\label{table_oscillations} \\
\hline \hline Class & Value $=\sum_i a_i$ & $\sum_i |a_i|$ & $\max_i |a_i|$ & $\sum_i \int |I_i(\underline{z})| d\underline{z}$ & $N_{\text{graph}}$ \\ \hline  \endfirsthead
\hline \endfoot  \hline \hline \endlastfoot
I(a) & 0.00047105(16) & $0.00047$ & $0.00047$ & $0.021$ & 1 \\
I(b1) & 0.0046715(11) & $0.0047$ & $0.0027$ & $0.12$ & 2 \\
I(b2) & 0.00233656(54) & $0.0023$ & $0.0014$ & $0.061$ & 2 \\
I(c) & 0.0234643(26) & $0.023$ & $0.012$ & $0.23$ & 3 \\
I(d) & 0.00380370(61) & $0.0038$ & $0.0033$ & $0.051$ & 2 \\
I(e) & 0.010289(11) & $0.074$ & $0.022$ & $0.76$ & 9 \\
I(f) & 0.00757106(47) & $0.0076$ & $0.0073$ & $0.037$ & 2 \\
I(g) & 0.0285696(18) & $0.029$ & $0.015$ & $0.15$ & 4 \\
I(h) & 0.0016826(63) & $0.13$ & $0.036$ & $0.63$ & 11 \\
I(i) & 0.01726(29) & $1.5$ & $0.12$ & $5.8$ & 39 \\
I(j) & 0.0004038(63) & $0.12$ & $0.061$ & $0.67$ & 2 \\
II(ap1) & 0.00924758(84) & $0.011$ & $0.0073$ & $0.067$ & 3 \\
II(ap2) & -0.0297767(11) & $0.030$ & $0.024$ & $0.11$ & 2 \\
II(ap3) & -0.0262183(11) & $0.026$ & $0.022$ & $0.096$ & 2 \\
II(as1) & -0.0466111(41) & $0.047$ & $0.031$ & $0.28$ & 2 \\
II(as2) & 0.0137916(15) & $0.017$ & $0.013$ & $0.15$ & 3 \\
II(as3) & -0.0299276(18) & $0.030$ & $0.023$ & $0.17$ & 2 \\
II(bp1) & 0.0326223(31) & $0.033$ & $0.011$ & $0.23$ & 6 \\
II(bp2) & -0.0800129(67) & $0.080$ & $0.045$ & $0.36$ & 4 \\
II(bp3) & -0.0836017(85) & $0.084$ & $0.045$ & $0.36$ & 4 \\
II(bs1) & -0.125057(11) & $0.13$ & $0.063$ & $0.86$ & 4 \\
II(bs2) & -0.125056(14) & $0.13$ & $0.062$ & $0.86$ & 4 \\
II(bs3) & 0.088110(10) & $0.093$ & $0.050$ & $0.96$ & 8 \\
II(bs4) & -0.180629(14) & $0.18$ & $0.048$ & $1.1$ & 8 \\
II(c1) & -0.0869389(86) & $0.087$ & $0.063$ & $0.42$ & 4 \\
II(c2) & 0.0375573(46) & $0.041$ & $0.024$ & $0.25$ & 6 \\
II(c3) & -0.0671244(72) & $0.067$ & $0.054$ & $0.29$ & 4 \\
II(d1) & -0.18988(11) & $1.6$ & $0.43$ & $6.8$ & 18 \\
II(d2) & 0.111341(75) & $0.87$ & $0.16$ & $4.3$ & 25 \\
II(d3) & -0.164378(83) & $1.3$ & $0.37$ & $4.9$ & 18 \\
II(e1) & -1.13757(46) & $20.5$ & $2.0$ & $65.0$ & 30 \\
II(e2) & -0.20478(28) & $24.2$ & $3.0$ & $38.5$ & 20 \\
II(f1) & -0.239896(88) & $0.32$ & $0.27$ & $2.5$ & 3 \\
II(f2) & -1.91510(23) & $3.8$ & $0.99$ & $14.9$ & 9 \\
II(f3) & -0.28054(18) & $8.6$ & $2.8$ & $10.3$ & 6 \\
III(ap1) & 0.056161(42) & $0.51$ & $0.16$ & $1.4$ & 10 \\
III(ap2) & 0.251184(26) & $0.45$ & $0.14$ & $0.89$ & 12 \\
III(ap3) & 0.227894(47) & $1.0$ & $0.12$ & $2.5$ & 20 \\
III(ap4) & 0.143139(17) & $0.36$ & $0.14$ & $0.64$ & 4 \\
III(ap5) & 0.071170(37) & $0.83$ & $0.15$ & $2.1$ & 20 \\
III(ap6) & 0.164795(20) & $0.37$ & $0.067$ & $1.1$ & 10 \\
III(ap7) & 0.0275451(27) & $0.075$ & $0.051$ & $0.15$ & 3 \\
III(as1) & 0.054599(55) & $1.4$ & $0.47$ & $2.6$ & 10 \\
III(as2) & 0.225730(17) & $0.47$ & $0.35$ & $0.79$ & 3 \\
III(as3) & 0.055588(66) & $2.4$ & $0.51$ & $4.9$ & 20 \\
III(as4) & 0.289427(35) & $0.78$ & $0.20$ & $1.7$ & 12 \\
III(as5) & 0.440024(55) & $1.2$ & $0.13$ & $4.0$ & 20 \\
III(as6) & 0.102979(21) & $0.33$ & $0.12$ & $1.1$ & 4 \\
III(as7) & 0.017022(35) & $0.45$ & $0.088$ & $2.0$ & 10 \\
III(b1) & 0.35850(13) & $4.0$ & $0.76$ & $10.7$ & 20 \\
III(b2) & 0.550730(74) & $1.7$ & $0.74$ & $3.4$ & 6 \\
III(b3) & -0.21825(16) & $8.9$ & $0.78$ & $22.4$ & 40 \\
III(b4) & 0.916024(91) & $3.3$ & $0.50$ & $8.1$ & 24 \\
III(b5) & 0.43819(10) & $2.8$ & $0.48$ & $6.3$ & 8 \\
III(b6) & 1.35323(15) & $5.1$ & $0.25$ & $16.7$ & 40 \\
III(b7) & -0.07113(10) & $2.8$ & $0.31$ & $10.1$ & 20 \\
III(c1) & 1.86018(48) & $13.0$ & $2.7$ & $37.4$ & 12 \\
III(c2) & 3.28050(91) & $43.3$ & $5.8$ & $104.7$ & 27 \\
III(c3) & 0.27166(48) & $14.3$ & $3.8$ & $32.7$ & 6 \\
III(c4) & -0.75943(79) & $33.1$ & $3.7$ & $89.8$ & 24 \\
III(c5) & 0.22308(51) & $36.5$ & $6.6$ & $66.4$ & 15 \\
III(c6) & 0.04392(46) & $31.4$ & $4.8$ & $50.3$ & 18 \\
IV(1) & -0.51586(43) & $15.7$ & $1.0$ & $29.9$ & 74 \\
IV(2) & -0.64878(28) & $7.1$ & $0.86$ & $12.5$ & 20 \\
IV(3) & -1.14824(60) & $30.9$ & $1.5$ & $63.8$ & 148 \\
IV(4) & 1.19593(38) & $15.1$ & $1.3$ & $27.5$ & 55 \\
IV(5) & -1.52785(60) & $32.4$ & $1.3$ & $67.4$ & 162 \\
IV(6) & 0.50531(30) & $8.4$ & $0.59$ & $17.4$ & 56 \\
IV(7) & -0.19295(26) & $4.6$ & $0.46$ & $11.0$ & 20 \\
IV(8) & -0.78444(38) & $10.0$ & $0.63$ & $25.6$ & 40 \\
IV(9) & -4.49029(73) & $33.4$ & $0.64$ & $84.4$ & 222 \\
IV(10) & 0.19698(32) & $8.3$ & $0.66$ & $21.9$ & 50 \\
IV(11) & 0.05233(55) & $18.7$ & $0.48$ & $60.7$ & 148 \\
IV(12) & -0.37250(31) & $6.6$ & $0.43$ & $19.0$ & 54 \\
V(1) & 6.169(20) & $599.3$ & $6.9$ & $1208.7$ & 706 \\
V(2) & 0.964(31) & $2077.2$ & $46.2$ & $2454.3$ & 148 \\
V(3) & 0.334(20) & $737.9$ & $38.0$ & $861.2$ & 55 \\
V(4) & -0.789(37) & $2029.2$ & $33.3$ & $3126.1$ & 706 \\
V(5) & -2.157(34) & $2253.3$ & $58.7$ & $3060.7$ & 370 \\
V(6) & -0.415(35) & $1948.0$ & $43.6$ & $2780.9$ & 558 \\
V(7) & 2.644(23) & $1062.3$ & $51.1$ & $1493.4$ & 261 \\
V(8) & -0.975(21) & $1028.0$ & $43.0$ & $1460.4$ & 336 \\
V(9) & 1.0807(63) & $137.5$ & $17.0$ & $200.4$ & 73 \\
VI(ap) & 0.482955(59) & $0.75$ & $0.49$ & $2.9$ & 5 \\
VI(as) & 0.558582(57) & $1.3$ & $0.70$ & $3.9$ & 5 \\
VI(b) & 1.34697(11) & $2.9$ & $1.1$ & $15.4$ & 10 \\
VI(c) & -2.53312(49) & $39.5$ & $3.9$ & $72.0$ & 37 \\
VI(d) & 1.8468(22) & $352.6$ & $12.1$ & $703.1$ & 127 \\
VI(e) & -0.43129(17) & $4.5$ & $1.1$ & $11.5$ & 13 \\
VI(f) & 0.77154(23) & $9.8$ & $0.63$ & $33.7$ & 46 \\
VI(g) & -1.5965(10) & $137.8$ & $3.8$ & $341.1$ & 122 \\
VI(h) & 0.18554(68) & $84.0$ & $2.2$ & $239.3$ & 162 \\
VI(i) & -0.04396(10) & $2.9$ & $0.49$ & $10.0$ & 16 \\
VI(j) & -0.22920(43) & $16.3$ & $4.7$ & $60.0$ & 9 \\
VI(k) & 0.67974(39) & $56.2$ & $5.1$ & $75.4$ & 32 \\
with lepton loops & -0.9377(35) & $1174.7$ & $12.1$ & $2656.6$ & 2323 \\
no lepton loops & 6.857(81) & $11872.6$ & $58.7$ & $16646.1$ & 3213 \\
\end{longtable}

\section{CONCLUSION}

The universal QED expansion coefficient $A_1^{(10)}$ of the lepton magnetic moments was  successfully calculated and compared with the previously known value obtained by T. Aoyama, M. Hayakawa, T. Kinoshita, M. Nio. The results agree very well in 31 out of 32 classes, but remain inconsistent in the remaining class. The deviating class value agrees with the value presented by the author in 2019, but not with the AHKN one.

Another calculations of $A_1^{(10)}$ are either not accurate enough to resolve the discrepancy or affecting only a very small part of the contributions.

The results presented are divided into 95 gauge-invariant classes, which subdivide the 32 classes mentioned above. The results at this level of detail were presented for the first time.

The author's method for removing divergences under the integral sign in Feynman parametric space was used. The method yields a finite integral for each Feynman graph. However, the  contributions of individual graphs remain large and oscillating relative to the final value. Changing the subtraction point to the off-shell one reduces the oscillations slightly, but does not solve the problem completely. These oscillations occur regardless of the subtraction method used and still require a comprehensive explanation.

A nonadaptive Monte Carlo integration algorithm was used for the numerical integration. This nonadaptivity means that the probability density function for each integral is predefined and remains fixed during the integration. However, this algorithm has 42 real parameters that can be chosen arbitrarily. 32 of them were adjusted during the experiments with the 4-loop calculations. 6 of them were chosen after poor convergence was observed for Set I(i) and some adjustments were made to make the convergence for this set as good as possible. Reasonable values were chosen for the remaining 4 parameters.

The definition of the probability density functions used is based on the combinatorics of the Feynman graphs. The procedure suitable for all Feynman graphs, including those with lepton loops, was presented for the first time. The description looks cumbersome at first glance, but it is accompanied by a number of examples that provide the opportunity to check and improve the algorithm.

\begin{acknowledgments}
The author would like to thank Gudrun Heinrich, Savely Karshenboim, Andrey Arbuzov, Lidia Kalinovskaya for important help. All preliminary calculations were performed using the ITP/TTP KIT computing cluster and  ITP KIT office computers. The main work was carried out on the HoreKa supercomputer funded by the Ministry of Science, Research and the Arts Baden-W\"{u}rttemberg and by the Federal Ministry of Education and Research. All images were drawn using AxoDraw and JaxoDraw. The research is supported by the Alexander von Humboldt Foundation. 
\end{acknowledgments}

\bibliography{amm5_a1_all}

\end{document}